\documentclass[manuscript,screen]{acmart}
\usepackage{soul}
\usepackage{xcolor}
\usepackage{booktabs}
\sethlcolor{yellow}
\usepackage{enumitem}
\usepackage{graphicx}
\usepackage{amsmath}
\usepackage{pgfplotstable}
\usepackage{listings}
\usepackage{xcolor} 
\usepackage{soul} 
\AtBeginDocument{%
  }

\renewcommand\footnotetextcopyrightpermission[1]{} 
\setcopyright{none} 
\acmVolume{}
\acmNumber{}
\acmArticle{}

\definecolor{codebg}{RGB}{245,245,245}  

\begin{document}

\renewcommand{\keywordsname}{Keywords}

\title{Beyond What Is Visible: Context-Aware Sentiment Across Online Communities Among Crisis Events}



\author{Zihui Ma}
\authornote{These authors contributed equally to this work.}
\authornote{Corresponding authors.}
\affiliation{%
 \institution{Emory University/University of Maryland}
 \city{Atlanta}
 \country{USA}}
\email{zihui.ma@emory.edu}

\author{Yiheng Chen}
\authornotemark[1]
\affiliation{%
 \institution{University of Alabama}
 \city{Tuscaloosa}
 \country{USA}}
\email{ychen226@crimson.ua.edu}

\author{Lingyao Li}
\affiliation{%
 \institution{University of Arizona}
 \city{Tucson}
 \country{USA}}
\email{lingyaoli@arizona.edu}

\author{Chen Chen}
\affiliation{%
 \institution{Florida International University}
 \city{Miami}
 \country{USA}}
\email{chechen@fiu.edu}

\author{Fan Yang}
\affiliation{%
 \institution{University of South Carolina}
 \city{Columbia}
 \country{USA}}
\email{YANG259@mailbox.sc.edu}

\author{Xiao Huang}
\affiliation{%
 \institution{Emory University}
 \city{Atlanta}
 \country{USA}}
\email{xiao.huang2@emory.edu}

\author{Haozheng Liu}
\affiliation{%
 \institution{New York University}
 \city{New York}
 \country{USA}}
\email{hl6602@nyu.edu}

\author{Yuhan Wang}
\affiliation{%
 \institution{University of Southern California}
 \city{Los Angeles}
 \country{USA}}
\email{ywang204@usc.edu}

\author{Runlong Yu}
\authornotemark[2]
\affiliation{%
 \institution{University of Alabama}
 \city{Tuscaloosa}
 \country{USA}}
\email{ryu5@ua.edu}

\author{Gregory B. Baecher}
\affiliation{%
 \institution{University of Maryland}
 \city{College Park}
 \country{USA}}
\email{gbaecher@umd.edu}


\renewcommand{\shortauthors}{Zihui Ma, Yiheng Chen, et al.}

\begin{teaserfigure}
    \centering
    \includegraphics[width=\textwidth]{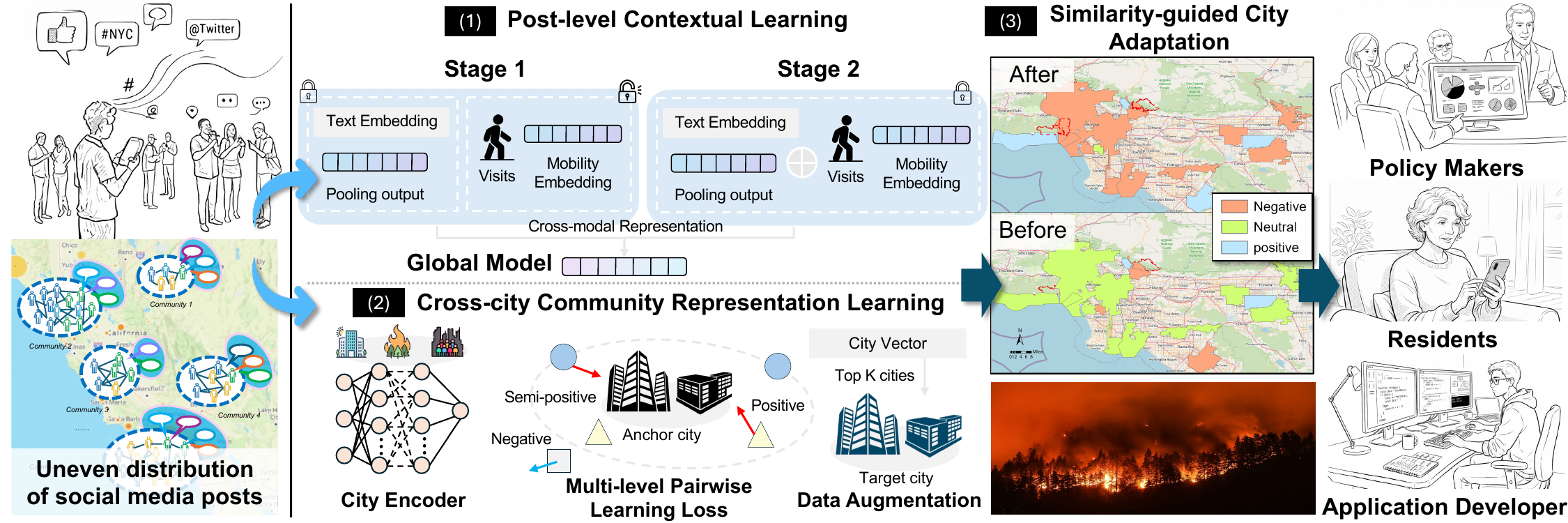}
    \caption{
       The proposed research framework that links what people say online, the community context they say it in, and how they move on the ground:  (1) post-level learning grounds text in mobility, (2) cross-city community learning relates communities by socioeconomic and hazard context, and (3) similarity-guided city adaptation lets data-sparse communities borrow evidence from structurally similar ones.}
    \label{fig: framework}
\end{teaserfigure}

\begin{abstract}
Social media offers timely insight into disaster experiences, yet online communities may not represent those affected on the ground. Uneven participation can bias crisis information systems toward data-rich locations and obscure communities with fewer digital traces. We address this with an adaptive cross-community learning framework that combines text and aggregate mobility signals with socioeconomic and hazard-informed community representations, supplementing sparse local observations with evidence from structurally related cities. We evaluate the framework on the January 2025 Southern California wildfires and 2024 Hurricane Helene in Florida. The adaptive model achieves the highest F1-score in both cases (0.5923 and 0.7615), while augmentation has its strongest effect where original observations are limited. The framework also reveals finer cross-city community-level sentiment variation and stronger sentiment-mobility correspondence than text-only models, with relationships differing across disaster contexts. These findings support crisis systems that better account for context-dependent evidence and improve interpretation of underrepresented community signals.
\end{abstract}

\maketitle

\section{Introduction}

Online social media provides a new lens for understanding how people experience disasters and for supporting disaster response. People use platforms such as X (formerly Twitter) to report local conditions, seek assistance, express concerns, and respond to evolving risks, making social media an important source of situational awareness \cite{ma2024surveying,dong2021social,li2023environmental}. However, what is available online is not necessarily representative of what affected communities experience. Social media users constitute a self-selected population, and unequal digital access and participation leave some communities better represented than others \cite{cesare2019understanding,dargin2021vulnerable,chen2022investigation}. Disaster-related social media attention often concentrates in populous areas \cite{fan2020spatial}, while vulnerable and digitally disconnected populations remain underrepresented \cite{samuels2020deepening,ramakrishnan2022examining,zhang2021revealing}. Such disparities can bias computational understanding toward digitally visible communities and overlook those with fewer online traces. Addressing this digital divide is therefore essential for building inclusive and context-aware crisis information systems.

Recent advances in natural language processing have substantially improved the ability to analyze social media text. Transformer-based models such as BERT \cite{vaswani2017attention, devlin2019bert} and RoBERTa \cite{liu2019roberta} can capture contextual and semantic nuances, enabling fine-grained sentiment and topic classification. These capabilities support a more comprehensive understanding of crisis communication, public awareness, and emotional dynamics \cite{liu2021crisisbert, paul2023fine, egger2022topic, grootendorst2022bertopic}. However, language models inherit biases from socially and geographically imbalanced training data, often amplifying existing inequities\cite{zheng2024reducing, weidinger2021ethical, navigli2023biases,gallegos2024bias}. Large language models (LLMs) further extend reasoning capacity but introduce additional limitations, including high computational cost, sensitivity to prompting, and limited transparency in commercial models \cite{zhao2023survey, yang2025comprehensive,li2025llms,yu2025survey}. As a result, sentiment models intended to reflect public opinion may inadvertently reinforce structural disparities. These limitations raise a central question: \textbf{\textit{how can computational models provide reliable and locally representative estimates of community sentiment when online evidence is uneven across affected places?}}

To address this challenge, we propose an \textbf{adaptive cross-city community learning framework} that contextualizes online sentiment using both community structure and aggregate behavioral evidence. The framework operates at two levels. The first layer models inter-city similarity using socioeconomic and demographic attributes to perform data augmentation for underrepresented regions. This process transfers knowledge from data-rich to data-sparse areas, improving generalization and fairness~\cite{luo2025great}.
Rather than assuming equivalence across communities, these relationships serve as a weak structural prior that allows data-sparse locations to draw on evidence from structurally related cities while preserving locally observed signals~\cite{yu2025physics}. The second layer integrates aggregate mobility with textual representations to provide behavioral context for interpreting online expression. By linking \textit{what people say, where they are, and how they respond}, the framework connects linguistic and behavioral evidence~\cite{li2025llms} while reducing dependence on the volume and composition of locally available posts. This provides a basis for examining how crisis systems can interpret sparse and uneven evidence across online communities.
%

Overall, our key contributions are twofolds:
\textbf{(1)} a novel adaptive learning framework that combines cross-city community representation learning with multimodal contextualization for disaster sentiment analysis under uneven online participation;
and \textbf{(2)} cross-disaster evidence showing how contextual grounding changes the representation and interpretation of community-level sentiment, with implications for designing crisis information systems that make sparse, transferred, and context-dependent evidence more visible to users.

\section{Related Work}
\subsection{Digital Inequity in Crisis Communication}

Crisis communication is an interactive and iterative process of exchanging information among stakeholders throughout the phases of mitigation, preparedness, response, and recovery \cite{reynolds2005crisis,haddow2013disaster,ma2024surveying}. Effective communication enhances public preparedness and reduces disaster risks, serving as a cornerstone of successful emergency management \cite{bradley2016effectiveness,savoia2013communications}. In recent years, social media platforms such as X, Facebook, and Instagram have become key venues for individuals to share their preparedness, perceptions, and experiences of disasters, accelerating the online circulation of crisis information. These crowd-sourced data have been increasingly leveraged to analyze public responses \cite{dong2021social,ma2024investigating,li2020leveraging}, including rescue coordination \cite{li2023environmental}, sentiment expression \cite{karimiziarani2025natural,gupta2025sentimentmapper,babu2025advanced}, and disaster information dissemination \cite{fan2025dynamics,ma2024investigating,al2025information}.

However, concerns about representativeness persist, as access to and engagement with digital platforms remain uneven across social and demographic groups. Social media users constitute a self-selected population rather than a random sample of society \cite{cesare2019understanding,chen2022investigation,dargin2021vulnerable}. Prior studies highlight that marginalized or digitally disconnected populations are systematically underrepresented in online disaster discourse \cite{ma2024surveying,fan2020spatial,wang2025mining}. For example, \citet{chen2022investigation} found that White users are overrepresented in disaster-related Twitter discussions, whereas Hispanic and Black populations are significantly underrepresented, with disparities varying by region. Similarly, \citet{mavrodieva2021social} observed that elderly populations face significant barriers to digital participation. Such imbalances can shape which experiences become visible to computational systems and which communities remain weakly represented in online crisis data. This makes representativeness a central challenge for social media–based disaster analysis, particularly when computational outputs are used to characterize community-level conditions or support crisis interpretation.

\vspace{-0.2cm}
\subsection{Context-aware sentiment modeling}

Sentiment analysis has long been a cornerstone of understanding public discourse, evolving from lexicon and feature-based methods to deep contextual encoders and, more recently, LLMs \cite{Wankhade2022SentimentSurvey,zhao2023survey}. Transformer-based models such as BERT\cite{devlin2019bert} and RoBERTa \cite{liu2019roberta} capture nuanced linguistic cues and have become standard for short, informal social media posts, with domain-specific variants like BERTweet achieving state-of-the-art accuracy on Twitter data and crisis-related text classification \cite{Nguyen2020BERTweet,liu2021crisisbert,paul2023fine}. However, despite impressive overall performance, these models often struggle to generalize across domains, regions, or events—precisely the settings where crisis communication unfolds. Benchmarks such as CrisisBench \cite{Alam2021CrisisBench} and WILDS \cite{Koh2021WILDS} reveal that sentiment classifiers trained in one geography or event type degrade sharply when applied elsewhere. This instability highlights a fundamental challenge: sentiment models capture what people say but often lack the geographic, behavioral, and community context needed to interpret those expressions across disaster settings.

Bias and representational imbalance in language models offer a related perspective on this problem \cite{Mehrabi2021BiasFairnessSurvey,Bansal2022NLPBiasFairness}. In natural language processing, bias can arise from stereotyped associations in learned representations \cite{sun2019mitigating,Czarnowska2021QuantifyingSocialBiasesNLP}, unequal error rates across demographic groups \cite{Mehrabi2021BiasFairnessSurvey,Czarnowska2021QuantifyingSocialBiasesNLP}, and the underrepresentation of particular communities in training data \cite{cesare2019understanding,chen2022investigation,ramakrishnan2022examining}. As models scale to LLMs, such biases may be reproduced or amplified through large and opaque pretraining corpora \cite{bender2021dangers,weidinger2021ethical,gallegos2024bias,navigli2023biases,Nadeem2021StereoSet}. Existing fairness-aware approaches address these issues through data augmentation \cite{Kaushik2020CDA}, representation-level debiasing \cite{Ravfogel2020INLP,Kaneko2021DebiasingContextualized}, and robust optimization \cite{Sagawa2020DRO}. While these approaches provide important mechanisms for improving group-level fairness, they typically focus on biases within observed training data and do not directly address the spatial and behavioral context underlying uneven participation in disaster-related social media.

Recent work in crisis informatics further shows that social media participation during disasters is geographically and socially uneven. Urban and higher-income communities tend to be more strongly represented, while rural and vulnerable populations may remain underrepresented in digital traces \cite{fan2020spatial,samuels2020deepening,zhang2021revealing,mavrodieva2021social,dargin2021vulnerable}. Text-only sentiment models may therefore produce community-level estimates that disproportionately reflect populations with greater online visibility. Addressing this gap requires moving beyond text-only modeling toward contextual approaches that combine online expression with evidence that is less dependent on social media participation itself. Mobility patterns, socioeconomic characteristics, and hazard exposure provide complementary signals for contextualizing online sentiment while reducing reliance on the volume and composition of locally available posts.

\section{Study Context and Data}
\label{sec:data}

\subsection{Disaster Cases}

By pairing a broad wildfire event with a regional hurricane, the following two cases provide contrasting settings for examining how uneven digital visibility, community context, and behavioral signals shape disaster sentiment inference.

\paragraph{2025 Southern California Wildfires.}
Our primary case study focuses on the January 2025 wildfires in Southern California, a region characterized by substantial urban--rural heterogeneity and recurring exposure to wildfire hazards. The fires collectively led to the evacuation of more than 200,000 residents, destroyed over 18,000 structures, and burned more than 57,000 acres \cite{calfire2025incidents}. Among the most severe events were the Eaton Fire in Altadena \cite{calfire2025eaton} and the Palisades Fire in Pacific Palisades \cite{caramela2025palisades}. Both fires affected densely populated wildland--urban interface (WUI) areas and produced substantial disruptions across socially and geographically diverse communities. The coexistence of urban, suburban, and less urbanized communities within the affected region provides a useful setting for examining geographically heterogeneous disaster experiences.

\paragraph{2024 Hurricane Helene.}
Our complementary case study examines Hurricane Helene in Florida, where coastal, metropolitan, and inland communities experienced substantially different levels of exposure and disruption. We analyze posts from September 20 to October 5, 2024, covering the period leading into Helene's development and landfall, the escalation of warnings and preparedness activities, and the immediate post-landfall period. Helene rapidly intensified over the Gulf of Mexico and made landfall in Florida’s Big Bend on September 26 as a Category 4 hurricane with maximum sustained winds of approximately 140 mph \cite{nhc2024helene}. The storm produced catastrophic surge in the Big Bend and significant coastal flooding farther south along Florida’s west coast, including the Tampa Bay region, where record-high water levels were observed at multiple tidal gauges \cite{wfts2024heleneclosures}. Impacts also extended well inland and across northern Florida. In Jacksonville, more than 224,000 customers lost power and hundreds of emergency road-access deployments were required, while Ocala experienced widespread tree damage and more than 45,000 customer outages \cite{nwsjax2024tornado}.

\subsection{Online Community Data}
We use X as the primary source of online community expression for both disaster cases. Posts were collected through the Brandwatch API \cite{brandwatch2025suite} using case-specific keyword-based queries that retrieved public posts matching disaster-related terms. For the California wildfire case, the query included ``wildfire'', ``bushfire'', ``California fire'', ``CAfire'', and ``CALfire''. For Hurricane Helene, the query combined event-specific terms such as ``hurricane helene'' and ``hurricane helen'' with broader hazard terms, including ``tropical storm,'' ``tropical cyclone,'' ``hurricane.'' Social media coverage varies substantially across communities. These posts therefore represent observations from users who were visible on the platform rather than a representative sample of all affected residents.

\subsection{Community Context Data}
\subsubsection{Aggregate Mobility}

We use mobility data as an aggregate behavioral signal that complements online expression with information about changes in population movement during disaster events. Mobility records for both case studies were obtained from the Veraset Visits dataset through the Dewey Data platform~\cite{veraset2022visits}. Veraset Visits is a device-level point-of-interest visitation dataset rather than a direct record of the movements of social media users. Using the same processing procedure for both cases, visit records were filtered by study period and affected state and converted into daily origin--destination (OD) flows at the Census tract level. OD pairs with fewer than five observed trips were excluded to reduce noise from sparse observations. The resulting flows characterize aggregate movement between geographic areas over the disaster period. These mobility data are not linked to individual social media users; they provide community-level behavioral context for interpreting online expression.


\subsubsection{Socioeconomic and Hazard Context}

We incorporate socioeconomic and hazard information to characterize the broader structural and environmental conditions of the study communities and to support representation learning. Socioeconomic attributes were obtained primarily from the American Community Survey (ACS) at the Census tract level and were combined with the urbanization measure used in our community feature set. The candidate variables describe demographic, economic, housing, infrastructural, and urbanization-related characteristics, providing context for comparing places with different structural conditions.

We also incorporate hazard-specific information from the FEMA National Risk Index~\cite{fema_nri}. For the California case, we retain the wildfire-specific risk variable used in our preprocessing pipeline; for the Florida case, we use the corresponding hurricane-specific risk variable. These variables capture relative geographic variation in hazard context. Socioeconomic and hazard attributes are subsequently harmonized to a common city-level representation, allowing the cross-community model to compare communities using both structural and hazard-related context.

\subsection{Data Preparation and Alignment}

For the California wildfire case, the raw collection contains 131,268 posts published between January 1 and January 31, 2025. We apply a common preprocessing pipeline to both disaster cases, including location inference, deduplication, data cleaning, restriction to observations with usable mobility coverage, and alignment with mobility observations from the same city and date. After preprocessing, the California case contains 28,267 unique posts, while the Hurricane Helene collection from September 20 to October 5, 2024 contains 40,880 unique posts. Posts with explicit geotags are prioritized for geographic assignment; when explicit geolocation is unavailable, locations are inferred from textual cues using an LLM \cite{chen2025empowering,yu2025rag} and, when necessary, supplemented with user-provided location metadata.

The tract-level mobility flows were then harmonized to the city scale used for the social media analysis. We used the National Historical Geographic Information System (NHGIS) Block Group-to-Places crosswalk \cite{ipums_nhgis_v20} to relate underlying Census geography to Census-defined places before combining flows into daily city-level mobility features. Socioeconomic and hazard data were harmonized to the same city-level reference. Socioeconomic variables were first screened using variance inflation factor (VIF) analysis to reduce multicollinearity and were then aggregated using population-weighted averaging; hazard variables were mapped through the same geographic framework.

The final dataset includes 29 California communities, comprising 13 urban and 16 non-urban locations, and 29 Florida communities, comprising 14 urban and 15 non-urban locations. Each community from one city is therefore represented through three complementary forms of evidence: social media posts capturing observed online expression, aggregate mobility characterizing population movement, and socioeconomic and hazard attributes describing community context.

\section{Methods}

\subsection{Framework Overview}

Our framework supports disaster sentiment inference when locally available social media evidence varies across communities. As illustrated in Figure \ref{fig: framework}, it consists of three connected components. First, post-level contextual learning combines textual expression with aggregate mobility context to form a shared multimodal representation. Second, cross-city representation learning models structural relationships among communities using socioeconomic and hazard characteristics. Third, similarity-guided city adaptation uses these learned relationships to supplement sparse local training evidence with weighted examples from structurally related communities.

The framework follows a shared-to-local learning strategy. A common multimodal model is first trained across communities, after which cross-city representations identify relevant sources of transfer for each target community. Adaptation then refines the shared model using both locally observed and similarity-weighted transferred evidence. This design preserves local observations while allowing communities with limited online data to benefit from broader contextual information.


\subsection{Post-level Contextual Learning}
\label{sec:individual-layer}

At the post level, we model textual expression together with aggregate mobility associated with the post’s geographic and temporal context. The goal is not to infer the behavior of an individual social media user, since the mobility data are aggregated and not linked to individual accounts. Instead, mobility provides community-level behavioral context that can help distinguish similar linguistic expressions occurring under different local conditions. We use a Transformer-based encoder-only architecture to represent post text. This architecture provides bidirectional contextual representations while remaining computationally efficient for the short social media posts in our dataset, which average approximately 16 tokens and contain fewer than 40 tokens in the longest sequence. In parallel, a feed-forward mobility encoder transforms the corresponding aggregate mobility features into a dense representation. The text and mobility embeddings are then combined through a fusion network before sentiment classification.

Formally, let $x^{(t)}$ and $x^{(m)}$ denote the textual and mobility inputs, respectively. A text encoder maps $x^{(t)}$ to contextual token representations $H = E_t(x^{(t)})\in\mathbb{R}^{L\times d}$, from which we derive a fixed-length post embedding by applying an affine transformation followed by $\tanh$ to the \texttt{[CLS]} representation:
\[
h_t = \tanh\!\big(W_p\,H_{[\texttt{CLS}]} + b_p\big).
\]

In parallel, a mobility encoder produces a mobility embedding $h_m = E_m(x^{(m)})$. The two embeddings are concatenated and transformed by a fusion network $F(\cdot)$ into a joint representation
\[
h_f = F([h_t; h_m]).
\]

A prediction head $C(\cdot)$ outputs sentiment probabilities $\hat{y} = C(h_f)$, and the model is trained with a standard cross-entropy loss. To construct a geographically aware base model, we apply a two-stage training strategy:

\emph{Stage 1: Weakly supervised pretraining}
We first train the multimodal classifier using a larger set of sentiment labels generated by \texttt{Mistral-8B-Instruct-2410}. We treat these labels as noisy supervision rather than unbiased ground truth. They provide large-scale initialization for learning coarse sentiment patterns and multimodal alignment. During this stage, we keep the text encoder frozen while updating the mobility encoder, fusion network, and classifier.

\emph{Stage 2: Human-annotated refinement.}
We continue training on a high-quality dataset labeled by human annotators. These reliable labels refine the decision boundary and reduce the influence of noise introduced by weak supervision. To calibrate the classifier without overwriting the learned multimodal alignment, both encoders are kept frozen and only the fusion network and classifier are updated. We denote the resulting model after Stage~2 by $f_\theta$ and refer to it as the global model. In the following subsections, we adapt $f_\theta$ to each city using the learned cross-city similarity structure.

\subsection{Cross-city Community Representation Learning}

Our cross-city community representation learning component models contextual relationships among online communities and borrows from recent augmentation-adaptive representation learning frameworks \cite{DBLP:journals/corr/abs-2509-14563}. The encoder uses socioeconomic and hazard-specific attributes as input, while hazard-risk proximity provides the ordering signal used to organize the learned representation space. Because the ranking objective does not directly supervise socioeconomic similarity, we interpret the resulting embeddings as risk-ordered contextual representations rather than evidence that two communities are equivalent.

Let $\mathcal{C}_{\mathrm{ctx}}$ denote the set of communities for which socioeconomic and hazard context is available, including cities that may not be represented in the social media data. Let $\mathcal{C}_{\mathrm{obs}} \subseteq \mathcal{C}_{\mathrm{ctx}}$ denote the subset of cities with labeled social media observations that can serve as adaptation targets or transfer sources. Each city $c \in \mathcal{C}_{\mathrm{ctx}}$ is represented by a static feature vector $f_c \in \mathbb{R}^{D}$. A city encoder $\phi: \mathbb{R}^{D} \rightarrow \mathbb{R}^{d}$ maps these features to a low-dimensional embedding $z_c = \phi(f_c)$. We adopt a \textbf{\emph{risk-aware ordinal ranking objective}} to learn graded relationships across cities rather than treating them as simply similar or dissimilar. For an anchor city $A$ and three candidate cities $P$, $S$, and $N$ (positive, semi-positive, and negative), selected according to differences in hazard-specific risk, the loss encourages risk-consistent relative similarity:

\begin{equation}
\begin{aligned}
\mathcal{L} = -\log \sigma\!\bigl(
& \lambda [\mathrm{sim}(z_A, z_P) - \mathrm{sim}(z_A, z_S)] + (1-\lambda)[\mathrm{sim}(z_A, z_S) - \mathrm{sim}(z_A, z_N)]
\bigr),
\end{aligned}
\end{equation}
where $\mathrm{sim}(\cdot,\cdot)$ denotes cosine similarity, $\sigma(u) = 1 / (1 + e^{-u})$ is the logistic sigmoid, and $\lambda \in [0,1]$ balances the two relative-similarity terms. The intended ordering is
\[
\mathrm{sim}(z_A,z_P) > \mathrm{sim}(z_A,z_S) > \mathrm{sim}(z_A,z_N).
\]
The current objective encourages this ordering through a weighted aggregate ranking signal; it does not impose the two inequalities as independent hard constraints.

For each mini-batch, anchor cities $A$ are uniformly sampled from $\mathcal{C}_{\mathrm{ctx}}$. Given the hazard-specific risk variable $r(c)$, other cities are grouped according to their absolute risk difference $|r(c)-r(A)|$ using thresholds $0 \le \tau_1 < \tau_2$:
\begin{equation}
\begin{aligned}
\mathcal{P}(A) &= \{c \in \mathcal{C}_{\mathrm{ctx}}\setminus\{A\} : |r(c)-r(A)| \le \tau_1 \}, \\
\mathcal{S}(A) &= \{c \in \mathcal{C}_{\mathrm{ctx}}\setminus\{A\} : \tau_1 < |r(c)-r(A)| \le \tau_2 \}, \\
\mathcal{N}(A) &= \{c \in \mathcal{C}_{\mathrm{ctx}}\setminus\{A\} : |r(c)-r(A)| > \tau_2 \}.
\end{aligned}
\end{equation}
Training tuples are constructed from communities available in the corresponding risk-difference pools so that the positive, semi-positive, and negative roles follow the ordering defined above.

After representation learning, we perform nearest-neighbor search in the learned embedding space for each target city $i \in \mathcal{C}_{\mathrm{obs}}$. Only communities with labeled social media observations are eligible to provide transferred training examples. We therefore define the candidate donor set as
\begin{equation}
\mathcal{M}(i) = \{\, j \in \mathcal{C}_{\mathrm{obs}} \setminus \{i\} \mid \mathrm{sim}(z_i, z_j) \ge 0 \,\},
\end{equation}
and define the Top-$K$ most similar communities as
\begin{equation}
\mathcal{N}_K(i) = \mathrm{TopK}_{j \in \mathcal{M}(i)}
\big(\mathrm{sim}(z_i,z_j)\big).
\end{equation}
We index these embeddings using FAISS for efficient retrieval. The resulting neighbor sets $\mathcal{N}_K(i)$ define the candidate sources of supplementary evidence used during community adaptation. Similarity is used to identify potentially relevant transfer sources and does not imply that two communities have equivalent experiences, needs, or sentiment distributions.

\subsection{Similarity-guided City Adaptation}

Similarity-guided adaptation supplements limited local evidence with weighted training examples from contextually related communities. Local observations retain full per-sample weight, while transferred observations are down-weighted according to cross-city similarity. This design lets the model benefit from broader regional information without assuming related communities share identical sentiment patterns.

For each target city $i$, let $\mathcal{D}_i=\{(x,y)\}$ denote its labeled dataset, where $x$ represents the multimodal input and $y\in\{-1,0,1\}$ denotes the sentiment label. We augment $\mathcal{D}_i$ with training examples from the Top-$K$ structurally related communities $\mathcal{N}_K(i)$ identified through the cross-city representation space. Contributions from neighboring city $j$ are weighted according to their learned similarity to the target city:

\begin{equation}
\alpha_{ij}
=
\frac{
\exp\left(\gamma \, \mathrm{sim}(z_i,z_j)\right)
}{
\sum_{k\in\mathcal{N}_K(i)}
\exp\left(\gamma \, \mathrm{sim}(z_i,z_k)\right)
},
\qquad
j\in\mathcal{N}_K(i),
\end{equation}

where $\gamma$ controls the sharpness of the similarity distribution. Each transferred sample $(x,y)$ from community $j$ receives weight $w(x,y)=\alpha_{ij}$, while samples observed in the target city $i$ retain full weight, $w(x,y)=1$. This weighting scheme preserves the distinction between locally observed evidence and information introduced through cross-city transfer, while allowing more structurally similar communities to contribute more strongly during adaptation.

We partition the parameters of the shared multimodal model $f_{\theta}$ as
$\theta=\{\theta_t,\theta_m,\theta_f,\theta_c\}$,
corresponding to the text encoder, mobility encoder, fusion network, and classifier, respectively. During community adaptation, the text and mobility encoders remain frozen to preserve the shared multimodal representation learned during pretraining and human-annotated refinement. The fusion network and classification head are then updated using the weighted combination of local and transferred observations. The adaptation objective minimizes the weighted empirical risk:

\begin{equation}
\min_{\theta_f,\theta_c}\mathcal{L}
=
\frac{1}{Z}
\sum_{(x,y)\in\mathcal{D}_i}
\mathcal{L}_{CE}\bigl(f_{\theta}(x),y\bigr)
+
\frac{1}{Z}
\sum_{j\in\mathcal{N}_K(i)}
\sum_{(x,y)\in\mathcal{D}_j}
\alpha_{ij}
\mathcal{L}_{CE}\bigl(f_{\theta}(x),y\bigr),
\end{equation}

where $\mathcal{L}_{CE}$ denotes cross-entropy loss and, for the objective written above,
\[
Z = |\mathcal{D}_i| + \sum_{j\in\mathcal{N}_K(i)} \alpha_{ij}|\mathcal{D}_j|
\]
is the total effective sample weight. Because $\alpha_{ij}$ is applied to every transferred sample from city $j$, the total contribution of a donor community depends on both its similarity weight and the number of transferred samples used from that community.
The resulting model is adapted to each target city while retaining a shared representation learned across the broader study region. Communities with limited local observations can therefore draw selectively on evidence from contextually related locations while retaining a clear distinction between directly observed and transferred supervision. This mechanism supports more stable sentiment estimation across unevenly observed communities without treating transferred evidence as equivalent to direct local observation.

\subsection{Evaluation Setup}

We evaluate the proposed framework across both disaster cases to examine predictive performance, representational effects, and the contribution of contextual information under different disaster conditions. We compare seven model configurations. VADER serves as a lexicon-based baseline, while RoBERTa and Mistral represent transformer- and LLM-based sentiment models, respectively. We further evaluate four variants of our framework to isolate the effects of cross-city learning and mobility integration: Global-Pure-Text, which uses textual information and is trained across all cities; City-Specific-Pure-Text, which adapts the text-based model to individual cities; Global-Fusion, which integrates textual and mobility features without cross-city learning; and City-Specific-Fusion, which combines multimodal fusion with cross-city learning. Model performance is evaluated using accuracy and macro-F1, together with class-wise F1-scores for negative, neutral, and positive sentiment.

We complement the benchmark evaluation with analyses of representation, behavioral grounding, and model robustness. We first examine sentiment distributions before and after similarity-guided augmentation across urban and non-urban locations to assess how cross-city adaptation affects data representation (Section~5.1). We then evaluate how mobility information shapes learned sentiment representations and examine the correspondence between predicted sentiment and observed mobility patterns (Section~5.2). Finally, we conduct ablation analyses of cross-city representation learning, encoder-freezing strategies, and alternative approaches to incorporating mobility information to assess each component's contribution (Section~5.3).

\section{Results and Evaluations}
\subsection{System Performance Across Communities}

\subsubsection{Classification Performance}
We first evaluate whether the proposed system can provide reliable sentiment estimates without allowing crisis interpretation to be dominated by communities that are already highly visible online. Table~\ref{tab:overall_performance_cases} presents overall F1-scores across the two disaster cases. The City-Specific -Fusion (adaptive model hereafter) consistently achieves the highest performance (F1 = 0.5923 for the California wildfire, 0.7615 for the Florida hurricane), and incorporating mobility information yields a similar advantage in both cases, with Global-Fusion reaching 0.5549 and 0.7483, respectively. These consistent patterns suggest that behavioral information, particularly when combined with similarity-guided adaptation, improves sentiment classification across disaster contexts.

The magnitude of improvement, however, varies considerably between cases. In California, VADER performs particularly poorly (F1 = 0.2010), while RoBERTa and Mistral improve substantially (0.4444 and 0.5107). Global-Pure-Text reaches only 0.2778, rising to 0.3967 with similarity-guided city adaptation and 0.5549 with the addition of mobility information (Global-Fusion); the adaptive model ultimately reaches 0.5923. In Florida, models generally perform more strongly and the gaps are narrower. VADER reaches 0.5238, RoBERTa and Mistral reach 0.7234 and 0.7363, respectively, and Global-Pure-Text and the City-Specific-Pure-Text(pure-text model hereafter) reach 0.7298 and 0.7266, respectively. Global-Fusion (0.7483) and the adaptive model (0.7615) again perform best. Notably, cross-city community learning alone does not consistently improve the pure-text model, particularly in Florida, whereas the adaptive model achieves the highest performance in both cases, indicating that the benefit of local adaptation depends on what it is combined with. The overall gap between California and Florida further points to the role of disaster-specific data characteristics.

\begin{table}[htbp]
  \centering
  \caption{Overall performance of sentiment models across California and Florida cases.}
  \vspace{-10pt}
  \label{tab:overall_performance_cases}
  \small
  \setlength{\tabcolsep}{6pt}
  \begin{tabular}{lcccc}
    \toprule
    & \multicolumn{2}{c}{California Case} &
    \multicolumn{2}{c}{Florida Case} \\
    \cmidrule(lr){2-3}\cmidrule(lr){4-5}
    & Accuracy & F1 & Accuracy & F1 \\
    \midrule
    VADER & 0.1954 & 0.2010 & 0.5300 & 0.5238 \\
    RoBERTa & 0.6051 & 0.4444 & 0.7500 & 0.7234 \\
    Mistral & 0.6792 & 0.5107 & 0.7500 & 0.7363 \\
    Global-Pure-Text & 0.7143 & 0.2778 & 0.7540 & 0.7298 \\
    City-Specific-Pure-Text & 0.7237 & 0.3967 & 0.7480 & 0.7266 \\
    Global-Fusion & \underline{0.7547} & \underline{0.5549} & \underline{0.7680} & \underline{0.7483} \\
    City-Specific-Fusion & \textbf{0.7722} & \textbf{0.5923} & \textbf{0.7780} & \textbf{0.7615} \\
    \bottomrule
  \end{tabular}
\end{table}

Table~\ref{tab:classwise_f1} further examines whether these patterns hold across sentiment classes. The adaptive model performs particularly well on neutral and positive sentiment in both cases, reaching F1-scores of 0.4677 in California and 0.7034 in Florida for positive sentiment, compared with 0.2155 and 0.6780 for Mistral, and 0.8687 in California and 0.7927 in Florida for neutral sentiment. The class-wise results reveal meaningful differences between disaster contexts. In California, gains are uneven across classes. Mistral achieves the highest negative-class F1 (0.5383), ahead of Global-Fusion (0.4790) and the adaptive model (0.4404), while the adaptive model shows a much larger advantage for positive sentiment. In Florida, by contrast, the adaptive model leads across all three classes, reaching 0.7885 for negative, 0.7927 for neutral, and 0.7034 for positive sentiment, with negative sentiment already captured relatively well by most non-lexicon models (RoBERTa, Mistral, and Global-Fusion reach 0.7558, 0.7816, and 0.7820, respectively).

\begin{table*}[htbp]
  \centering
  \caption{Class-wise F1 score breakdown across California and Florida cases.}
  \label{tab:classwise_f1}
  \small
  \setlength{\tabcolsep}{5pt}
  \begin{tabular}{lcccccccc}
    \toprule
    & \multicolumn{4}{c}{\textit{California Case}}
    & \multicolumn{4}{c}{\textit{Florida Case}} \\
    \cmidrule(lr){2-5} \cmidrule(lr){6-9}
    Model
      & F1 & Neg\_F1 & Neu\_F1 & Pos\_F1
      & F1 & Neg\_F1 & Neu\_F1 & Pos\_F1 \\
    \midrule
    VADER
      & 0.2010 & 0.2424 & 0.1419 & 0.2186
      & 0.5238 & 0.6037 & 0.4926 & 0.4752 \\
    RoBERTa
      & 0.4444 & 0.4485 & 0.7202 & 0.1644
      & 0.7234 & 0.7558 & 0.7773 & 0.6370 \\
    Mistral
      & 0.5107 & \textbf{0.5383} & 0.7782 & 0.2155
      & 0.7363 & 0.7816 & 0.7494 & 0.6780 \\
    Global-Fusion
      & \underline{0.5549} & \underline{0.4790}
      & \underline{0.8524} & \underline{0.3333}
      & \underline{0.7483} & \underline{0.7820}
      & \underline{0.7818} & \underline{0.6812} \\
    City-Specific-Fusion
      & \textbf{0.5923} & 0.4404
      & \textbf{0.8687} & \textbf{0.4677}
      & \textbf{0.7615} & \textbf{0.7885}
      & \textbf{0.7927} & \textbf{0.7034} \\
    \bottomrule
  \end{tabular}
\end{table*}

\subsubsection{Urban–Non-Urban Representation}
\begin{figure}[htbp]
    \centering
    \includegraphics[width=0.9\textwidth]{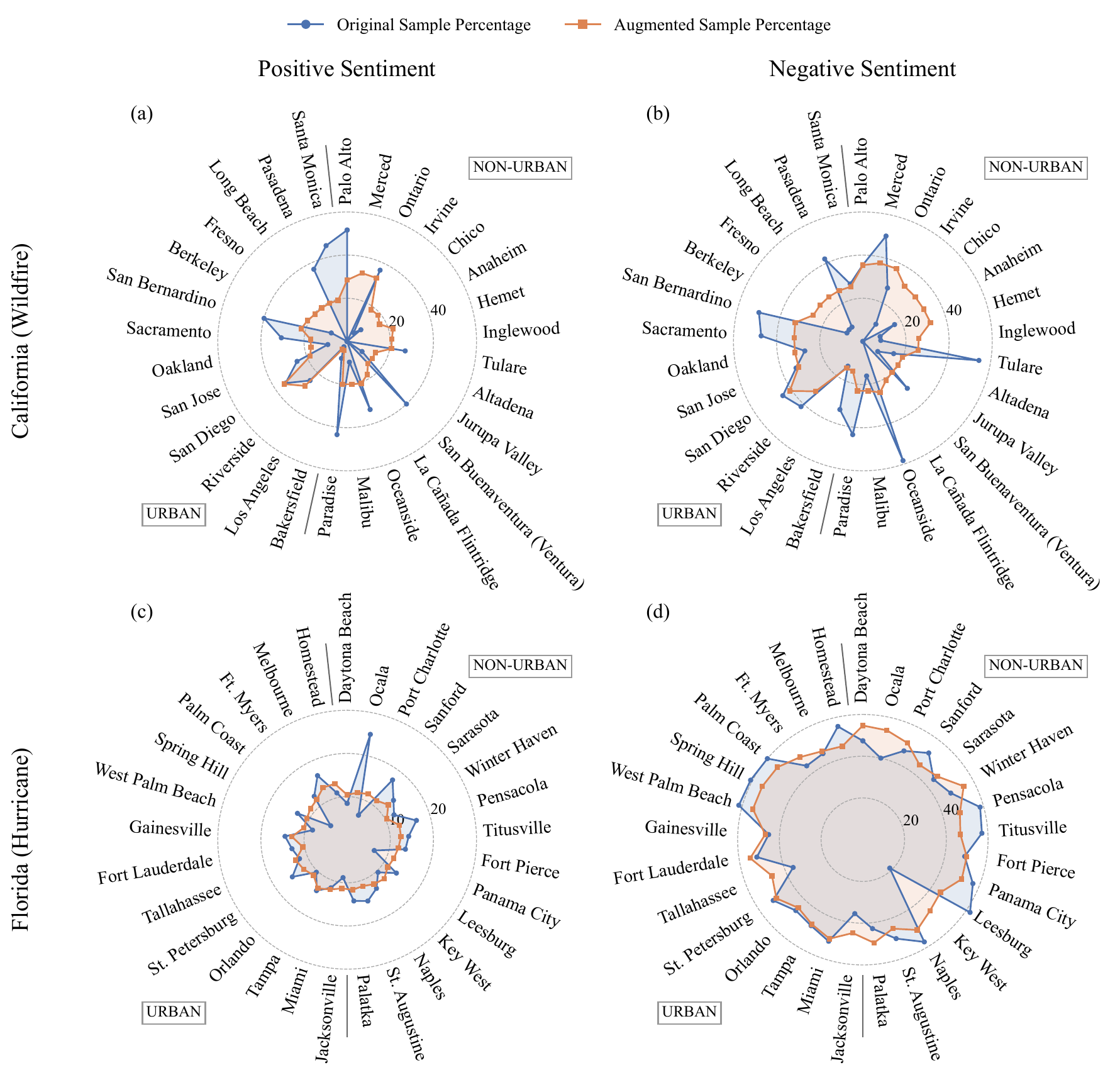}
    \vspace{-4pt}
    \caption{The distribution of community sentiment before (blue) and after (orange) augmentation, for positive (a,~c) and negative (b,~d) posts. Communities with little original online presence gain the most representation.}
    \vspace{-10pt}
    \label{fig:label}
\end{figure}

Beyond predictive performance, we examine whether similarity-guided augmentation changes representational coverage in communities with limited original online observations. We adopt urban and non-urban status as a proxy for digital visibility; this classification likely correlates with other factors, such as internet access and baseline posting volume, and we treat it here as a starting point for examining representational disparities rather than a complete explanation of their cause. As shown in Figure~\ref{fig:label}, we visualize the proportions of positive and negative sentiment before and after applying data augmentation. In the original sample distribution (denoted in blue), substantial class imbalance and high variance are observed, particularly among non-urban cities. In the California case, communities from cities such as La Ca\~{n}ada, Flintridge, and Chico exhibit no negative-sentiment tweets, with all three registering 0\% negative sentiment. The augmented distribution (denoted in orange) yields more balanced coverage across cities, an effect that is particularly pronounced in the negative sentiment class, where numerous non-urban locations attain clearer representation following augmentation.

By enriching data in under-sampled areas, the augmentation process reduces some of the representation gaps between urban and non-urban communities. In California, urban centers such as Riverside (positive rising from 25\% to 28.45\%, negative falling from 41.67\% to 31.9\%) and San Diego (positive rising from 34.48\% to 35.24\%, negative falling from 44.83\% to 40.95\%) already show relatively balanced sentiment distributions. Non-urban communities exhibit larger changes by comparison. In Tulare, positive and negative sentiment shift from 27.27\% and 54.55\% to 20.78\% and 25.97\%, respectively, while in Merced, positive sentiment rises from 0\% to 34\% and negative sentiment falls from 50\% to 39\%. These shifts indicate that augmentation has a larger effect in places where the original online observations provide a more limited view of local sentiment.

A comparable pattern emerges in the Florida hurricane case, although the effect of augmentation is less pronounced. Urban communities from cities such as Tampa (positive: 13.78\% to 13.15\%; negative: 48.08\% to 47.32\%) and Jacksonville (positive: 8.85\% to 11.36\%; negative: 35.63\% to 44.97\%) remain relatively stable following augmentation, whereas Palm Coast (positive: 5.00\% to 11.16\%; negative: 60.00\% to 53.65\%) and the non-urban city of Daytona Beach (positive: 8.42\% to 10.41\%; negative: 47.37\% to 54.69\%) exhibit more visible shifts. The negative sentiment distribution, however, remains closely aligned before and after augmentation across most cities, consistent with the class-wise results reported above, which indicate that most models already capture negative sentiment well in Florida. A possible explanation is that hurricane-related observations already convey a comparatively strong and consistent negative signal, leaving less room for augmentation to reshape the distribution. From a system-level perspective, these results indicate that the proposed framework does not uniformly alter community signals; rather, it introduces additional representation where digital visibility is more limited.

\subsection{Localized Sentiment and Behavioral Response}

\begin{figure*}[htbp]
    \centering
    \includegraphics[width=0.9\textwidth]{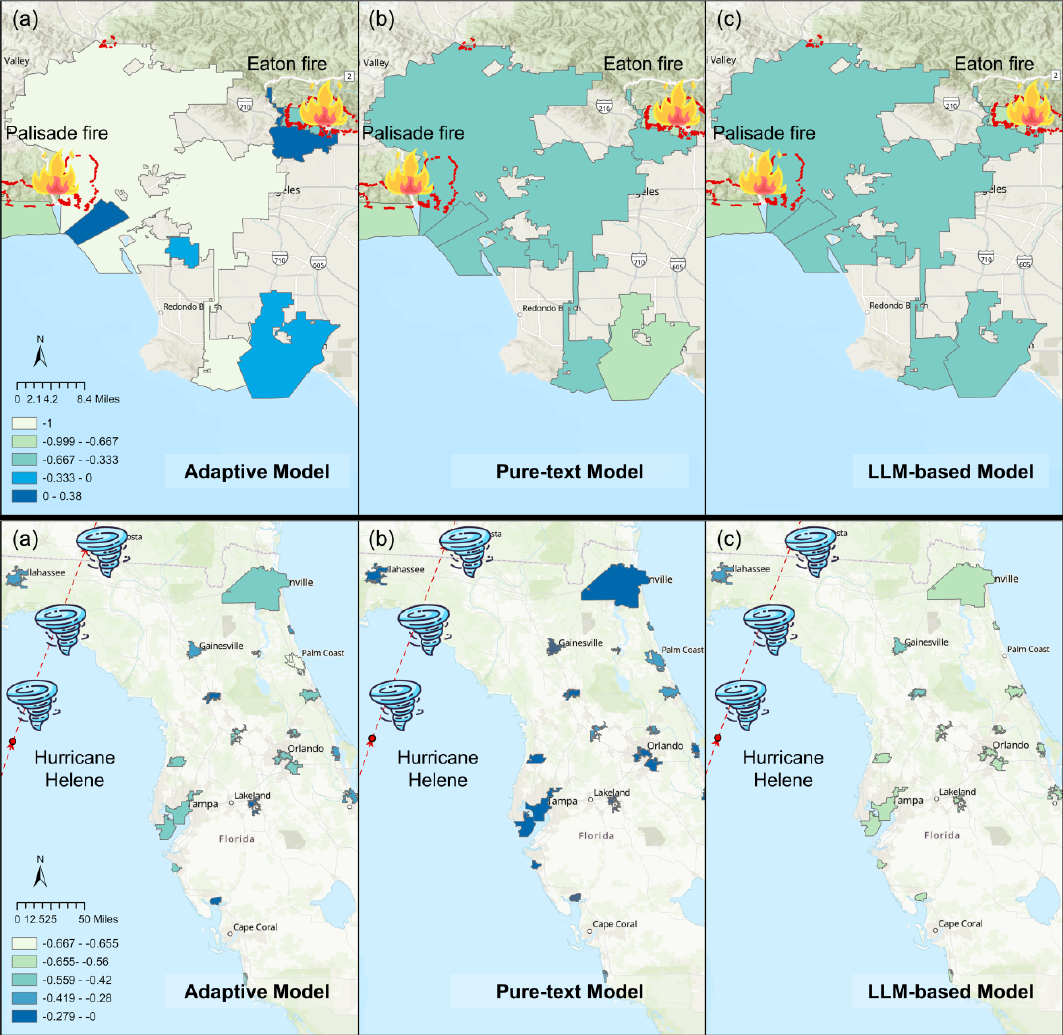}
    \caption{Community-level sentiment distribution from (a) the adaptive, (b) pure-text, and (c) LLM-based models under the same cross-city learning setting. The pure-text and LLM estimates blur neighboring communities together; the adaptive model shows local differences, e.g., between Pasadena and Altadena.}
    \vspace{-5pt}
    \label{fig: sentiment}
\end{figure*}

\subsubsection{Spatial Sentiment Variation}

To further assess the system's ability to capture localized differences in social media users' experiences and responses to disasters, we examine public sentiment trends over time and space. We introduce the metric of \textit{accumulative sentiment}, which smooths temporal fluctuations and highlights overall emotional responses to wildfire events. Formally, the accumulative sentiment score at time $t$ is defined as:
\begin{equation}
\text{Acc\_sentiment}(t) = 
\frac{\sum_{i=1}^{t} (\text{pos}_i - \text{neg}_i)}{\sum_{i=1}^{t} (\text{pos}_i + \text{neg}_i)}
\end{equation}
where $\text{pos}_i$ and $\text{neg}_i$ represent the number of positive and negative tweets, respectively. This metric captures the evolving emotional tone over time, smoothing short-term volatility from low tweet volumes \cite{li2020leveraging}. Using this metric, we visualized city-level sentiment distributions produced by three models: our proposed adaptive model, a fine-tuned text-only transformer (Pure-text) model, and an LLM-based model on Mistral (Figure \ref{fig: sentiment}).

In the California case, the adaptive model produces more geographically diverse sentiment estimates. In the broader Los Angeles city, sentiment drops to $-1$ under the adaptive model, compared to $-0.5$ (Pure-text) and $-0.33$ (LLM). In coastal areas such as Malibu, sentiment estimates range from $-0.58$ (LLM) to $-0.70$ (adaptive). For cities near the Eaton fire, sentiment improves under the adaptive model, with Pasadena shifting from $-0.61$ (LLM) and $-0.47$ (Pure-text) to $+0.25$ (adaptive), and Altadena from $-0.65$ (Pure-text and LLM) to $-0.40$ (adaptive). Urban–suburban sentiment gradients are also more pronounced. For example, Santa Monica maintains a relatively neutral tone ($0.21$), while adjacent Malibu shows more negativity ($-0.70$) in the adaptive model. Pasadena and Altadena show differing sentiment directions despite their proximity. These distinctions are less visible in the other models, highlighting the adaptive approach's greater spatial sensitivity.

On the other hand, Pure-text and the LLM-based model capture a broadly negative regional response in Florida case, but their estimates remain comparatively uniform across locations (sample-size-weighted standard deviations of 6.66 and 6.11 percentage points, respectively). The adaptive model preserves this overall negative tendency while introducing greater city-level variation (a sample-size-weighted standard deviation of 7.22 percentage points), yielding a more differentiated representation of local sentiment. Rather than substantially altering the dominant regional interpretation, the adaptive system refines it by rendering smaller differences among communities more visible. This granularity matters because communities affected by the same regional disaster may still express differing levels of concern, disruption, or recovery, and a uniform regional signal risks obscuring that variation.

\subsubsection{Sentiment-Mobility Relationship}
Understanding the interplay between public sentiment and physical mobility is critical during disasters, as it reveals how emotional responses influence real-world behavior \cite{xu2025impact, li2020leveraging}. To explore this relationship, we focus specifically on the first layer of our framework, post-level contextual learning, which integrates both sentiment and mobility signals into our framework. We begin with a t-SNE--based visualization to assess how including mobility embeddings affects sentiment representation, comparing the adaptive model (with mobility) with the pure-text model (without mobility).

\begin{figure}[htbp]
    \centering
    \includegraphics[width=\textwidth]{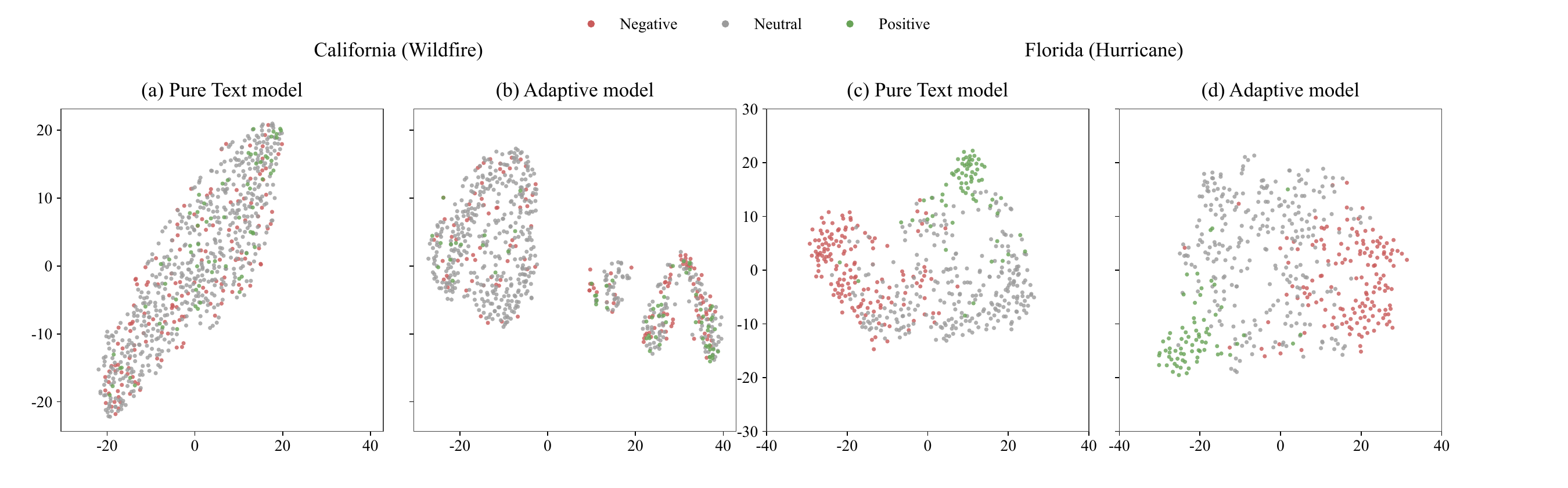}
    \caption{Effect of mobility embeddings on sentiment representations via t-SNE after post-level contextual learning: California wildfire case (left two panels) and Florida hurricane case (right two panels).}
    \label{fig:tsne}
\end{figure}

As shown in Figure~\ref{fig:tsne}, mobility information reorganizes the representation space in both cases, though in distinct ways. In the wildfire case, the pure-text model forms a largely continuous representation in which negative, neutral, and positive observations are highly intermixed. With mobility information, the adaptive model splits this continuous space into several more distinct local regions, although sentiment labels remain mixed within each region. Rather than simply improving class separation, this pattern suggests that behavioral context exposes additional heterogeneity among users whose textual expressions might otherwise appear similar. In the hurricane case, the effect is more continuous. The adaptive representation does not fragment into distinct local regions; rather, sentiment observations become more systematically differentiated across the space while retaining substantial overlap. 

\begin{figure*}[htbp]
    \centering
    \includegraphics[width=\textwidth]{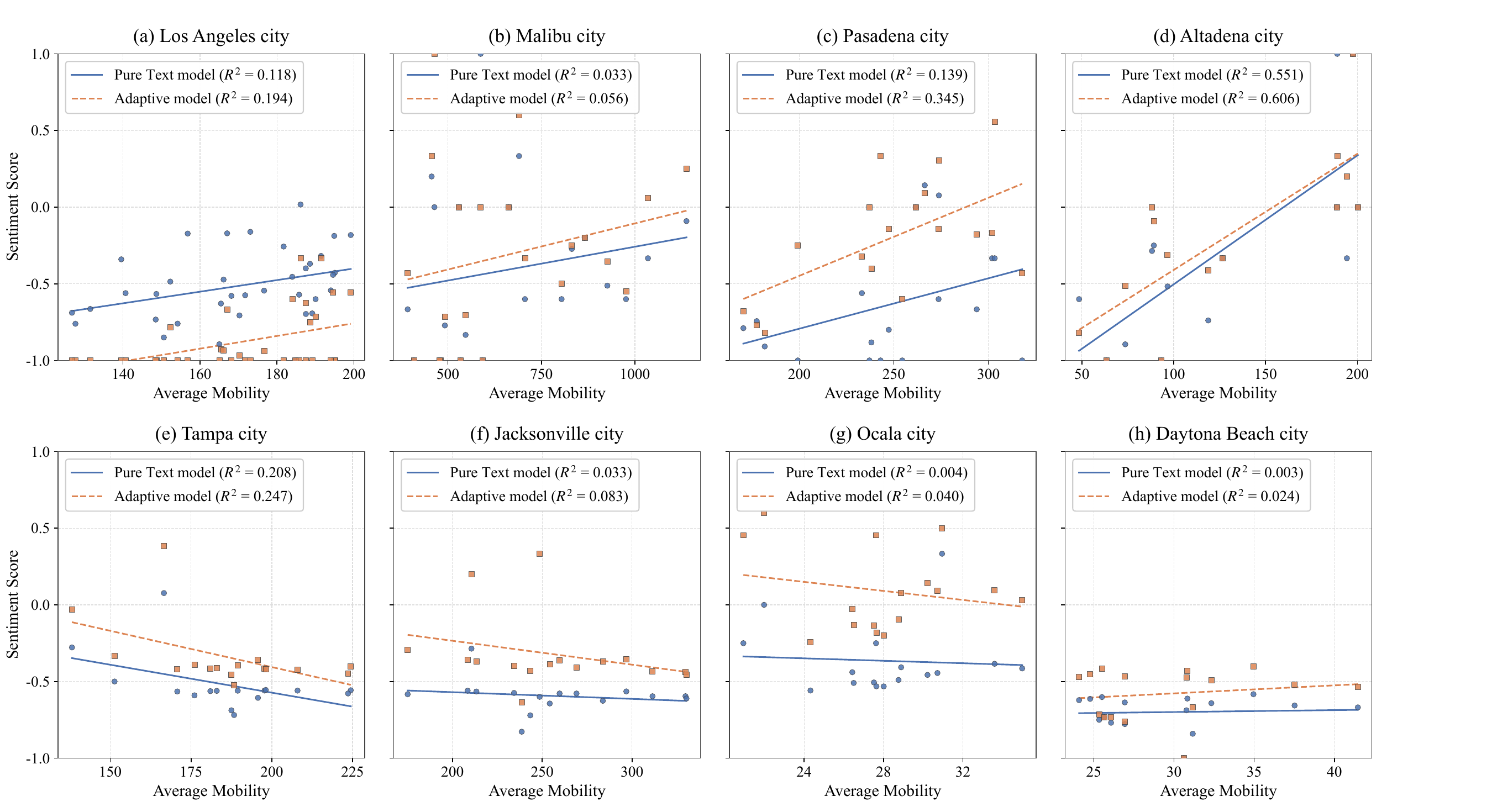}
    \vspace{-20pt}
    \caption{The correlation between online sentiment and physical movement in selected communities (blue: text-only; orange: adaptive). The relationship differs by crisis context: positive in the wildfire, negative in the hurricane.}
    \vspace{-10pt}
    \label{fig:regression}
\end{figure*}

To further quantify this relationship, we conduct a correlation analysis across four representative communities in each disaster case, as shown in Figure~\ref{fig:regression}. For California, Los Angeles and Malibu represent communities affected by the Palisades Fire, while Pasadena and Altadena represent communities affected by the Eaton Fire, with each pair capturing urban and non-urban contexts. For Florida, Tampa, Jacksonville, Ocala, and Daytona Beach were selected as representative communities distributed along the  path that were affected by Hurricane Helene. In each subplot, the orange dots and regression lines represent correlations derived from our adaptive model, while the blue ones correspond to the Global-Pure-Text model. 

Overall, we observe consistently positive correlations between average sentiment and mobility in California, indicating that periods of more positive online expression tend to align with increased physical movement. The adaptive model shows stronger and more stable correlations than the pure-text model, most notably in suburban Altadena ($R^2 = 0.606$ vs.~0.551). In Florida, by contrast, the relationship is predominantly negative across Tampa, Jacksonville, and Ocala, while Daytona Beach  (\mbox{$R^2 = 0.024$ vs.\ $0.003$}) shows only a weak association. Although the improvements in urban Los Angeles and  Tampa are relatively modest, the general pattern suggests that incorporating mobility signals enables the model to better capture co-occurring behavioral shifts, providing a more contextually grounded understanding of community dynamics during crises. 

\subsection{Robustness and Ablation Analysis}
To evaluate the impact of our architectural and training design choices, we conduct a set of ablation studies focusing on city-level representation, training strategies, and multimodal input handling. 

\begin{figure}[htbp]
    \centering
    \includegraphics[width=\textwidth]{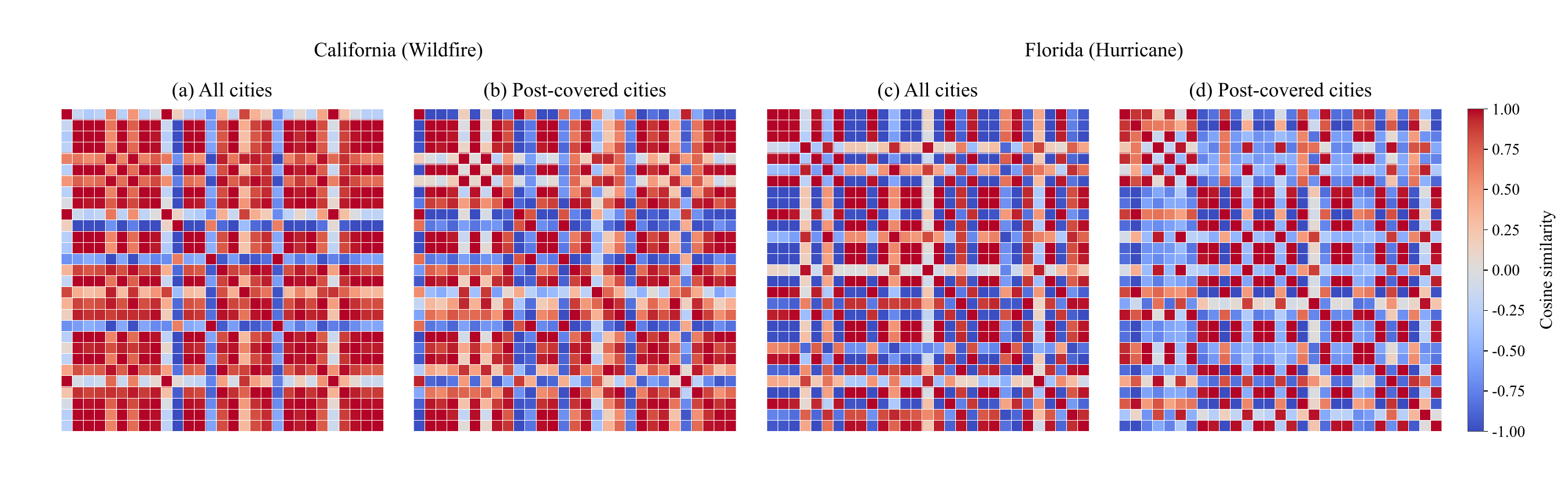}
    \vspace{-10pt}
    \caption{Community similarity learned from all cities (a,~c) versus only the cities visible on social media (b,~d). Learning beyond the digitally visible cities keeps community relationships from depending on who happened to post.}
    \vspace{-10pt}
    \label{fig:A2SL}
\end{figure}


\subsubsection{Generalization of cross-tity representations}
We first examine the role of broader city representation learning in shaping stable city embeddings. Figure~\ref{fig:A2SL} visualizes the cosine similarity matrices for city representations under two training configurations: one using all 938 California cities available in the socioeconomic dataset (left), and the other using all 956 Florida cities available in the corresponding dataset (right). When representation learning uses the full set of cities, the similarity structure exhibits smoother gradients and more coherent cluster blocks, with higher within-cluster similarity and clearer separation across city groups. In contrast, restricting training to post-covered cities produces a noisier and less interpretable structure, characterized by more fragmented similarity patterns. These results suggest that incorporating cities beyond those directly observed in the social media data provides a broader structural context for learning city representations and supports more stable cross-city relationships. 

Similarly, in the Florida case (Figure~\ref{fig:A2SL}), learning from the broader set of available locations (29 cities) produces a more coherent similarity structure than restricting training to cities covered by social media data. This consistency across cases suggests that city-wide learning reduces the dependence of the representation space on which communities happen to be digitally visible during a disaster.

\subsubsection{Training strategy for model stability}
\begin{figure}[htbp]
    \centering
    \includegraphics[width=0.85\textwidth]{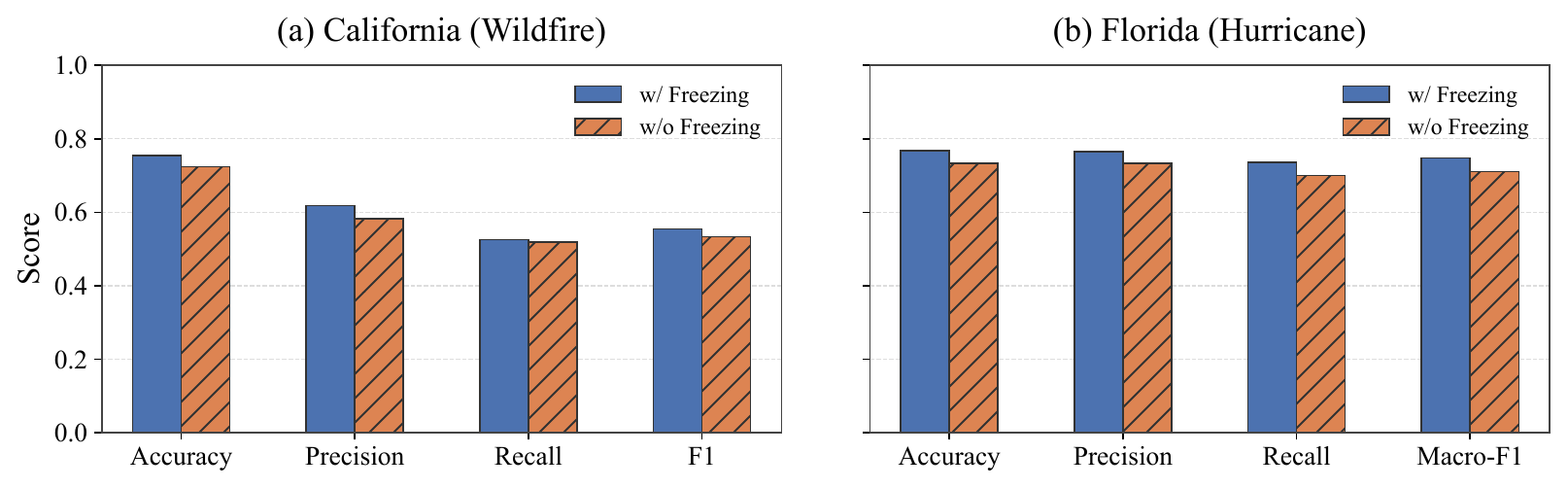}
    \caption{Training strategy comparison. Adaptation with frozen (blue) versus unfrozen (orange) encoders. }
    \vspace{-10pt}
    \label{fig:freeze}
\end{figure}

We then compare two supervised refinement strategies to assess how different levels of encoder flexibility influence learning. Based on Figure~\ref{fig:freeze}, freezing both the text and mobility encoders while training only the perception layer consistently yields better performance across all metrics. 
This finding reinforces that, under limited training data, fine-tuning only the final layer preserves general representations while reducing the risk of overfitting. Conversely, fully unfreezing the encoders leads to diminished scores, suggesting that excessive parameter updates introduce instability during the adaptation phase. 

\subsubsection{Effectiveness of multimodal fusion.}
Finally, Table~\ref{tab:Ablation} presents a comparison of two input designs for incorporating mobility signals: a prompt-based approach that verbalizes mobility as textual input, and a multimodal fusion strategy that treats mobility as a separate numerical modality. Across both global and city-specific settings, fusion-based models consistently outperform their prompt-based counterparts, with F1-scores improving from $0.35$ to $0.56$ (wildfire case) and $0.69$ to $0.76$ (hurricane case) in the city-specific configuration. These results underscore that treating mobility as structured input rather than auxiliary text allows the model to capture richer behavioral signals and integrate them more effectively into sentiment inference.

\begin{table*}[t]
  \centering
  \caption{Ablation study of different mobility input strategies.}
  \small
  \setlength{\tabcolsep}{7pt}
  \begin{tabular}{lcccccc}
    \toprule
    & \multicolumn{3}{c}{California Case}
    & \multicolumn{3}{c}{Florida Case} \\
    \cmidrule(lr){2-4} \cmidrule(lr){5-7}
    Model & Accuracy & Recall & F1 & Accuracy & Recall & F1 \\
    \midrule
    Global-Prompt
      & 0.7224 & 0.3575 & 0.3281
      & 0.7300 & 0.6838 & 0.7006 \\
    City-Specific-Prompt
      & 0.7412 & 0.4097 & 0.4184
      & 0.7140 & 0.6711 & 0.6885 \\
    Global-Fusion
      & \underline{0.7547} & \underline{0.5252} & \underline{0.5549}
      & \underline{0.7680} & \underline{0.7367} & \underline{0.7483} \\
    City-Specific-Fusion
      & \textbf{0.7722} & \textbf{0.5652} & \textbf{0.5923}
      & \textbf{0.7780} & \textbf{0.7556} & \textbf{0.7615} \\
    \bottomrule
  \end{tabular}
  \label{tab:Ablation}
\end{table*}

\section{Discussion}

Our findings suggest that social-media-based disaster sensing is shaped not only by \textit{what} people express online, but by \textit{where} those expressions originate and \textit{how} communities respond behaviorally. These patterns recurred across both the California wildfire and Florida hurricane cases, suggesting they are not artifacts of a single disaster type. Performance differences between global and locally adapted models, together with variation in sentiment-mobility relationships across communities, indicate that disaster information systems should not treat online content as uniformly representative of ground truth. We discuss these implications along three dimensions: supporting community expression, designing representation-aware crisis interfaces, and informing emergency decision-making.

\subsection{Supporting Context-Rich Crisis Communication} 

Disaster communication is an interactive process among stakeholders who collectively make sense of critical situations \cite{reynolds2005crisis, haddow2013disaster}. Crisis-related posts are often brief because authors and their intended audiences share situational knowledge that need not be restated. Prior HCI research similarly shows that terse social media posts can remain meaningful during disasters when readers already understand the surrounding event context \cite{vieweg2010microblogging}. Computational systems, however, do not have access to this shared context. In California, supplying socioeconomic, hazard, and mobility information substantially improves classification performance, whereas the gain is smaller in Florida, where negative hurricane-related sentiment is already well captured by text alone. Together, these cases suggest that the value of contextual grounding depends on how much ambiguity remains in the original text. Crisis information systems should therefore be evaluated not only on the language models they deploy, but also on the contextual evidence they can assemble to support more meaningful communication between affected communities and those responding to them.

Contextual grounding also reveals local differences that text-only analysis can miss. Pasadena and Altadena, both affected by the same wildfire, appear relatively similar under text-only estimates, whereas our context-aware model separates them substantially, consistent with documented differences in evacuation timing \cite{beckett2025lafires}. In Florida, the pure-text and LLM-based models already capture a broadly negative regional response but produce comparatively uniform estimates across locations; the adaptive model preserves this regional signal while revealing finer city-level variation. Context is therefore useful both for recovering distinctions from sparse or ambiguous observations and for preventing dominant regional patterns from obscuring local variation. In the wildfire case, contextual information is especially beneficial for the positive class, where brief expressions of relief, gratitude, and reassurance may depend heavily on circumstances left unstated in the post. Making these localized differences more visible can support crisis communication that better reflects changing conditions across communities, helping residents interpret evolving situations and navigate available information more effectively \cite{ma2024surveying, li2020leveraging}.

Where contextual information enters the system also matters. We tested mobility information both as verbalized text and as a separate structured signal, and the latter performed substantially better. From an HCI perspective, this suggests that crisis systems should not place the burden of contextualization on affected users. Requesting additional location tags, categories, or explanations during emergencies adds interaction costs precisely when users and responders alike face limited attention and capacity \cite{10.1145/3613904.3642368}. Systems should instead assemble available context around how people already communicate, so that residents can continue sharing information naturally while the system provides richer contextual support for interpretation. This matters most for underrepresented communities, since keyword-based collection and language models alike can privilege already common vocabularies and expressions \cite{gallegos2024bias, navigli2023biases, cesare2019understanding, chen2022investigation, dargin2021vulnerable, mavrodieva2021social}.

Our framework draws contextual evidence from sources that require no additional participation from users, including ACS community characteristics, FEMA hazard information, and aggregated mobility data. The city-representation ablation supports this design: learning community relationships across the full set of 938 study locations in California yields a more coherent structure than restricting learning to the 29 locations with observed posts. A similar pattern is observed in Florida when learning is expanded to all 956 study locations. This cross-disaster consistency suggests that broader contextual evidence can help compensate for uneven digital participation without requiring less-visible communities to contribute more data, consistent with structured transfer approaches in low-resource settings \cite{luo2025great}

\subsection{Representation-Aware Interface Design} 

Our findings suggest that crisis interfaces should make data representation visible with model outputs. A sentiment map alone can make a city with little online activity appear equivalent to one supported by hundreds or thousands of posts. Interfaces should therefore show indicators such as post volume, location coverage, and class balance together with city-level estimates. For example, a map might use color to show sentiment while indicating how much local evidence supports each estimate and whether a city is underrepresented relative to its population. This matters because disaster-related social media attention tends to concentrate in populous areas, while gaps in online visibility are often socially patterned~\cite{chen2022investigation, ma2024surveying, wang2025mining}. It also builds on HCI work showing that geospatial visualizations should support different forms of local knowledge and decision contexts rather than assume a single interpretation~\cite{saha2022visualizing}.

Interfaces should also distinguish \textit{observed} from \textit{inferred} information. In our framework, estimates for data-sparse cities are partly informed by similarity-weighted transfer from data-richer cities. Displaying these estimates identically could obscure their provenance and imply stronger local evidence than the data support. Interfaces could therefore use provenance labels and uncertainty cues to show which cities contributed information, which attributes determined similarity, and how much local evidence supports each estimate. When uncertainty can be quantified, ranges or graded cues may be more appropriate than a single precise value. Prior visualization research shows that hiding missing observations can bias interpretation and that uncertainty displays affect how people assess imputed data~\cite{sarma2022evaluating}; recent HCI work similarly shows how uncertainty from data quality and visualization choices can shape sensemaking~\cite{dritsa2024design}. These cues can frame transferred estimates as signals that require additional local evidence rather than substitutes for community reports.

In addition, sentiment and mobility should be presented as coordinated but distinct signals. Our results show that their relationship varies across cities and across disaster types, with sentiment-mobility correspondence strengthening under the adaptive model in both cases but to differing degrees, so combining them into a single ``community condition'' score could obscure meaningful differences. Instead, a coordinated view could align sentiment trends, mobility changes, hazard exposure, and key events over time, allowing users to see where these signals converge or diverge. For example, negative sentiment with little mobility change may reflect constrained movement, delayed information, or limited data. The interface should surface these possibilities rather than imply a single explanation.

Last, interfaces can support contextual interpretation through filters for time, geography, source, and data availability. Privacy-preserving access to representative posts can also help users examine the evidence behind an estimate. Interfaces should allow comparisons between raw and augmented distributions and show which cities contributed to transferred estimates. As a result, these features shift crisis interfaces away from sentiment scores toward a more auditable view of what is observed, inferred, and missing. These recommendations remain design implications rather than evaluated features. Future work could co-design and test them with emergency personnel and affected communities, particularly whether coverage and uncertainty cues reduce overconfidence without obscuring urgent patterns.

\subsection{Equitable Crisis Decision-Making } 

For emergency decision-makers, our framework adds an extra layer of evidence to identify communities whose online visibility may not fully reflect local conditions. Emergency-management agencies increasingly use social media to support situational awareness and public communication, yet practitioners still face limitations in staffing, verification capacity, and uneven information coverage \cite{dargin2021vulnerable, fan2020spatial}. Our framework addresses one part of this challenge by combining cross-city adaptation with mobility-informed behavioral context, allowing locations with sparse or potentially misleading online signals to become more visible within the decision process. Rather than treating sentiment as a direct measure of community need, these outputs indicate where additional local verification may be warranted and where sentiment and mobility should be interpreted jointly rather than as a single fixed signal. In this sense, the framework is intended to direct attention and verification rather than automate resource allocation.

Both disaster cases illustrate this role, though in complementary ways. In California, the adaptive model estimates cumulative sentiment of $-0.70$ for Malibu and $-0.40$ for Altadena, and Altadena shows the strongest sentiment-mobility correspondence among the selected communities ($R^2 = 0.606$ versus $0.551$ for Global-Pure-Text), suggesting that its online sentiment co-varies more closely with observed mobility than in the other selected communities. Los Angeles, by contrast, shows a comparatively weaker relationship ($R^2 = 0.194$ versus $0.118$), indicating that aggregate sentiment and mobility are less closely aligned in this large and heterogeneous urban setting. In Florida, the relationship changes direction rather than simply weakening: sentiment and mobility are negatively associated across Tampa, Jacksonville, and Ocala, even as the adaptive model produces a stronger correspondence than Global-Pure-Text at each location. These observations do not establish a direct link between sentiment, mobility, and specific response failures or community needs. Their value lies instead in identifying locations and disaster contexts where additional communication, field assessment, or local evidence may be particularly important before computational signals inform action.

HCI research on AI-assisted decision-making similarly emphasizes that effective decision-support systems should advance human reasoning rather than offer recommendations to be accepted or rejected \cite{sivaraman2026intelligent}. Our two cases suggest that the system can support this reasoning by highlighting locations where online expression and behavioral evidence show unusually strong, weak, or context-dependent relationships. These signals can then be interpreted alongside emergency-call records, shelter demand, infrastructure outages, traffic conditions, damage assessments, and reports from local organizations, rather than acted on in isolation. The contrast between California and Florida further shows that the meaning of these relationships is context-dependent: a positive sentiment-mobility association in one disaster and a negative association in another should not be interpreted through the same operational rule. Likewise, the comparatively weak correspondence observed in Los Angeles indicates that the relationship between online expression and behavioral response is less interpretable at the aggregate level, and that stronger corroborating evidence may therefore be needed.

Our system also has implications for how decision-makers engage with uncertainty. Human-AI interaction research shows that task uncertainty can substantially shape users' trust and reliance on AI outputs \cite{salimzadeh2024dealing}. Estimates supported by substantial local evidence should therefore not be interpreted in the same way as those relying more heavily on transferred information or sparse observations. Our representation results make this distinction concrete: in Merced, positive sentiment representation increases from $0\%$ to $34\%$ after augmentation, meaning the resulting distribution depends substantially on information transferred from structurally similar cities rather than on additional local posts. Decision interfaces can support more calibrated judgment by making evidence coverage, provenance, and uncertainty visible, and by allowing users to examine how an estimate was produced. Such transparency is particularly important for low-visibility communities that conventional social media monitoring might otherwise overlook. Evaluation should therefore extend beyond classification performance to examine whether these systems help decision-makers identify informational blind spots, improve investigation coverage across communities, and appropriately interpret uncertain or transferred evidence.

\subsection{Limitations and Future Work}
Several limitations suggest opportunities to extend this work. First, social media provides only a partial and uneven view of affected communities. Keyword-based retrieval may miss locally specific or evolving crisis language, while incomplete geographic information introduces uncertainty when locations must be inferred. Similarity-guided augmentation can supplement sparse local observations with evidence from structurally related cities, but it cannot recover the perspectives of residents who are absent from the platform or the collected data, nor does it make the resulting dataset representative of the full affected population. Cross-city transfer also depends on how well socioeconomic and hazard attributes capture meaningful relationships among cities, motivating the incorporation of additional local evidence, finer-grained contextual attributes, and alternative similarity measures to characterize communities with limited digital visibility better.

Second, mobility and sentiment remain incomplete signals of disaster experience. Mobility is aggregated and not linked to individual social media users, and observed movement may reflect routine travel, economic activity, or other contemporaneous factors in addition to disaster-related responses. Because mobility also feeds into the post-level contextual learning model, the observed sentiment-mobility correspondence should not be interpreted as independent behavioral validation. Sentiment labels are similarly reductive, since crisis-related posts may combine emotional expression, factual reporting, requests for assistance, and institutional communication. Incorporating complementary behavioral and situational signals, such as emergency calls, shelter demand, traffic conditions, or infrastructure disruptions, could provide a broader basis for examining how different forms of evidence jointly support crisis interpretation.

Finally, our analysis operates at the city level and across only two disaster cases, limiting both spatial granularity and generalizability. Neighborhood-level differences in exposure, socioeconomic conditions, infrastructure access, and digital participation may be obscured within city-level representations, motivating evaluation at finer geographic scales and across a broader range of hazards, regions, and temporal settings. The present study also evaluates the computational framework rather than direct user interaction with its outputs, so the interface and decision-support recommendations in Section 6 remain design implications rather than evaluated system features. Extending this work through co-design and user evaluation with emergency personnel and affected communities would help determine how to communicate transferred evidence, provenance, uncertainty, and differences between online expression and behavioral context in practice.

\section{Conclusion}

This paper addresses the challenge of generating reliable and locally representative disaster sentiment estimates when online evidence is unevenly distributed across affected communities and language models may reproduce biases embedded in socially and geographically imbalanced data. We develop an adaptive cross-city community learning framework that integrates textual expression with aggregate mobility context and augments sparse local observations through similarity-guided transfer based on socioeconomic and hazard characteristics. Evaluated across the Southern California wildfires and Hurricane Helene in Florida, the adaptive model achieves the strongest overall sentiment-classification performance, while augmentation yields the largest gains in representational coverage in communities with limited original observations. The framework also preserves finer-grained cross-city community-level sentiment variation than text-only and LLM-based approaches. In addition, it strengthens the correspondence between online sentiment and observed mobility patterns, although the direction of this relationship varies across disaster contexts. These findings suggest that contextual grounding can improve sentiment inference while reducing reliance on uneven digital participation and text-only representations that may carry forward existing geographic biases.

More broadly, this work argues that social media sentiment should not be treated as a direct proxy for community conditions, nor should language-model outputs be assumed to provide context-neutral representations of disaster experience. Crisis informatics systems become more informative when they make visible how much evidence is locally observed, how much is transferred from structurally related communities, and where linguistic and behavioral signals converge or diverge. Such transparency can help users identify informational blind spots, interpret model outputs with appropriate caution, and determine where additional local verification may be needed. By foregrounding representational uncertainty and community context, this work shifts disaster sentiment analysis from simply classifying online expression toward designing systems that support human interpretation of incomplete, uneven, and model-mediated digital evidence. It underscores the importance of representation-aware crisis systems that connect partial social media traces with broader community context and human judgment.

\bibliographystyle{ACM-Reference-Format}
\bibliography{main}

\clearpage

\section{Acknowledgment}
The authors gratefully acknowledge Fangqi Chen and Afra Izzati Kamili for their contributions to the manual annotation effort. We also thank Dr. Zhaoxi Zhang and Dr. Yuki Miura for their early structural feedback, which helped shape the initial development of the framework.

\section{Appendices}
\label{sec:appendix}

\subsection*{Ethics}
This study utilizes social media data obtained under an academic research license, subject to the platform’s data use agreement and ethical research guidelines. While the data are not publicly available, access was granted for non-commercial, scholarly purposes and complied with all applicable terms of service and institutional data governance policies. All records were anonymized and processed in aggregate to protect user privacy, and no personally identifiable information (PII) was collected, stored, or disclosed. Human annotations for sentiment labeling were performed by trained researchers with informed consent. The study adheres to ethical standards of digital research, emphasizing fairness, transparency, and responsible interpretation to avoid reinforcing social or spatial inequities in disaster-related analyses and decision-making.

\subsection*{Sentiment prompt}
\begin{lstlisting}[language=Python]
SENTIMENT_PROMPT = """
Task:
You are a careful annotator for social-media posts. Your job is to assign one sentiment label to each tweet: negative, neutral, or positive. Think through the decision silently. Do not reveal your reasoning. Output only the final JSON object.

Input:Tweet Text: {tweet}

Instructions:
Annotation policy (apply in this order):
1. Negative if the tweet conveys complaint, fear, loss, harm, danger, evacuation stress, property/health damage, anger, blame, or explicitly negative emotion--even if mixed with facts.
2. Positive if it clearly expresses relief, gratitude, appreciation, praise, successful containment/help, hopeful outlook or other explicit positive emotion.
3. Neutral if it's informational/ambiguous (e.g., resource provided such as a website/shelter link, questions without clear emotion, logistics, "only hashtags/keywords with no substantive content," or irony that cannot be resolved from context).
4. Emojis, intensifiers, negations, and sarcasm matter only when unambiguous from the text.
5. Hashtags and mentions count only if they clearly express sentiment (e.g., #grateful).
6. When unsure, choose neutral.

Few-shot examples:
<Positive tweet examples> 

<Negative tweet examples> 

<Neutral tweet examples> 

Output:
Output must be in strict JSON format with the following structure:
{{
    "sentiment": "<negative|neutral|positive>"
    "confidence": 0-1
}}

"""
\end{lstlisting}

\subsection*{Descriptive statistics of the variables.}

Guided by prior research \cite{ma2025analyzing} and the Social Vulnerability Index (SVI) framework—which spans socioeconomic status, household composition, racial and ethnic minority status, and housing and transportation—we first designated 22 tract-level ACS indicators as core controls, given their established importance for measuring social vulnerability and disaster risk. Building on this foundation, we assembled an extended pool of candidate predictors for the city-wide learning layer and applied a Variance Inflation Factor (VIF) screening procedure to the remaining 79 non-core variables (Table~\ref{tab:vif_final_features}).

\begin{table}[htbp]
\centering
\small
\caption{Final VIF values for the selected ACS tract-level features.}
\label{tab:vif_final_features}
\begin{tabular}{lrlr}
\toprule
Feature & Final VIF & Feature & Final VIF \\
\midrule
pct\_occupation\_healthcare\_support & 4.8787 & pct\_units\_2                 & 2.6711 \\
pct\_commute\_carpool                & 4.5577 & pct\_units\_1\_attached       & 2.6203 \\
pct\_occupation\_business            & 3.9466 & pct\_units\_1\_detached       & 2.5293 \\
pct\_married\_now                    & 3.7147 & pct\_occupation\_architecture & 2.2026 \\
pct\_rent\_1000\_1499                & 3.7038 & pct\_occupied\_housing        & 2.1364 \\
pct\_rent\_1500\_1999                & 3.6981 & pct\_occupation\_legal        & 1.9880 \\
pct\_complete\_kitchen               & 3.5558 & pct\_commute\_public\_transit & 1.9229 \\
pct\_occupation\_computer            & 3.5024 & median\_rooms                 & 1.8662 \\
pct\_occupation\_social\_service     & 3.4445 & pct\_occupation\_management   & 1.7596 \\
pct\_rent\_2000\_more                & 3.2875 & pct\_commute\_walk            & 1.6206 \\
pct\_occupation\_protective          & 3.1240 & pct\_rent\_under\_500         & 1.5962 \\
pct\_industry\_construction          & 3.0623 & pct\_complete\_plumbing       & 1.5530 \\
pct\_graduate\_degree                & 3.0438 & pct\_no\_telephone            & 1.5415 \\
pct\_occupation\_education           & 3.0158 & households\_cohabiting        & 1.5331 \\
pct\_households\_living\_alone       & 2.8306 & pct\_units\_3\_4              & 1.4679 \\
pct\_occupation\_food\_service       & 2.8292 & pct\_native\_alone            & 1.3940 \\
pct\_rent\_500\_999                  & 2.7753 & pct\_heating\_fuel\_none      & 1.3876 \\
occupants\_per\_room                 & 2.7522 & pct\_heating\_fuel\_oil       & 1.1258 \\
 &  & pct\_heating\_fuel\_wood & 1.0171 \\
\bottomrule
\end{tabular}
\end{table}

We iteratively removed the variable with the highest VIF until all surviving predictors had VIF values below the conventional threshold of 5, thereby mitigating multicollinearity and improving model stability. This process yielded 37 additional predictors, which we combined with the 22 core ACS indicators to obtain a final set of 59 tract-level variables for the city-wide learning layer. These variables serve as structured inputs for generating city embeddings in our adaptive model, enabling it to represent social and infrastructural heterogeneity across both urban and non-urban areas. Descriptive statistics for the tract-level variables are summarized in Table~\ref{tab:ca_tract_demographics_stats}.

\begin{table*}[htbp]
\centering
\caption{Summary statistics of census-tract-level demographic features in California.}
\label{tab:ca_tract_demographics_stats}
\resizebox{\textwidth}{!}{%
\begin{tabular}{@{}lrrrrr@{\hspace{1.2em}}lrrrrr@{}}
\toprule
Feature & $N$ & Mean & Std. & Min & Max & Feature & $N$ & Mean & Std. & Min & Max \\
\midrule
urban\_rate & 9097 & 92.69 & 23.18 & 0.00 & 100.00 & pct\_occupation\_education & 9056 & 5.56 & 3.90 & 0.00 & 100.00 \\
median\_household\_income & 8988 & 99732.01 & 46278.58 & 9417.00 & 250000+ & pct\_occupation\_healthcare\_support & 9056 & 3.77 & 3.14 & 0.00 & 32.66 \\
mean\_household\_income & 9029 & 127680.52 & 64137.98 & 13384.00 & 603656.00 & pct\_occupation\_protective & 9056 & 2.09 & 2.24 & 0.00 & 32.31 \\
pct\_black\_alone & 9080 & 5.60 & 8.50 & 0.00 & 90.60 & pct\_occupation\_food\_service & 9056 & 5.40 & 3.93 & 0.00 & 36.96 \\
pct\_hispanic\_latino & 9080 & 38.64 & 26.45 & 0.00 & 100.00 & pct\_commute\_carpool & 9056 & 9.53 & 6.11 & 0.00 & 100.00 \\
pct\_white\_not\_hispanic & 9080 & 36.90 & 25.14 & 0.00 & 100.00 & pct\_commute\_public\_transit & 9056 & 3.67 & 6.17 & 0.00 & 78.00 \\
pct\_asian\_alone & 9080 & 14.41 & 16.28 & 0.00 & 100.00 & pct\_commute\_walk & 9056 & 2.58 & 5.37 & 0.00 & 100.00 \\
fem\_pct & 9080 & 49.87 & 5.30 & 0.00 & 100.00 & pct\_complete\_plumbing & 9042 & 99.56 & 1.56 & 43.80 & 100.00 \\
male\_pct & 9080 & 50.13 & 5.30 & 0.00 & 100.00 & pct\_overcrowded & 9042 & 8.99 & 9.02 & 0.00 & 78.95 \\
pct\_hs\_grad\_higher & 9076 & 83.85 & 14.00 & 0.00 & 100.00 & pct\_no\_telephone & 9042 & 1.00 & 1.57 & 0.00 & 20.50 \\
less\_9th\_pct & 9076 & 9.09 & 9.53 & 0.00 & 78.20 & pct\_units\_1\_attached & 9044 & 7.01 & 8.53 & 0.00 & 85.70 \\
pct\_bachelor\_higher & 9076 & 35.30 & 21.79 & 0.00 & 100.00 & pct\_units\_3\_4 & 9044 & 5.38 & 6.59 & 0.00 & 72.10 \\
unemployment\_rate & 9056 & 6.58 & 4.21 & 0.00 & 66.70 & median\_rooms & 9029 & 5.10 & 1.15 & 1.30 & 9.0+ \\
poverty\_rate & 9049 & 12.57 & 9.73 & 0.00 & 100.00 & pct\_occupied\_housing & 9044 & 92.67 & 9.02 & 0.00 & 100.00 \\
eld\_65p\_pct & 9080 & 15.71 & 8.88 & 0.00 & 100.00 & pct\_industry\_construction & 9056 & 6.83 & 4.80 & 0.00 & 100.00 \\
youth\_18l\_pct & 9080 & 21.50 & 7.40 & 0.00 & 72.70 & pct\_rent\_under\_500 & 9028 & 3.39 & 6.98 & 0.00 & 95.80 \\
avg\_household\_size & 9039 & 2.96 & 0.69 & 1.04 & 7.19 & pct\_rent\_500\_999 & 9028 & 9.62 & 13.45 & 0.00 & 100.00 \\
pct\_owner\_occupied & 9042 & 55.77 & 24.77 & 0.00 & 100.00 & pct\_rent\_1000\_1499 & 9028 & 19.31 & 17.33 & 0.00 & 100.00 \\
pct\_foreign\_born & 9080 & 26.10 & 13.90 & 0.00 & 100.00 & pct\_rent\_1500\_1999 & 9028 & 21.72 & 16.22 & 0.00 & 100.00 \\
per\_capita\_income & 9076 & 46813.76 & 27489.78 & 482.00 & 278590.00 & pct\_rent\_2000\_more & 9028 & 45.95 & 30.73 & 0.00 & 100.00 \\
median\_age & 9078 & 38.76 & 7.96 & 10.80 & 82.30 & pct\_heating\_fuel\_oil & 9042 & 0.26 & 1.74 & 0.00 & 42.90 \\
total\_population & 9129 & 4311.11 & 1739.09 & 0.00 & 38907.00 & pct\_cohabiting\_households & 9042 & 7.37 & 4.78 & 0.00 & 76.92 \\
built\_2020\_plus\_pct & 9044 & 0.37 & 1.34 & 0.00 & 35.50 & pct\_households\_living\_alone & 9042 & 23.14 & 12.66 & 0.00 & 100.00 \\
pct\_native\_alone & 9080 & 1.03 & 1.98 & 0.00 & 73.20 & pct\_married\_now & 9080 & 46.20 & 11.96 & 0.00 & 100.00 \\
pct\_occupation\_management & 9056 & 11.14 & 6.73 & 0.00 & 100.00 & pct\_graduate\_degree & 9076 & 13.53 & 11.69 & 0.00 & 77.98 \\
pct\_occupation\_business & 9056 & 5.62 & 4.05 & 0.00 & 40.00 & pct\_units\_1\_detached & 9044 & 58.96 & 28.70 & 0.00 & 100.00 \\
pct\_occupation\_computer & 9056 & 3.75 & 4.71 & 0.00 & 51.84 & pct\_units\_2 & 9044 & 2.42 & 3.93 & 0.00 & 53.50 \\
pct\_occupation\_architecture & 9056 & 2.38 & 2.66 & 0.00 & 23.22 & pct\_complete\_kitchen & 9042 & 98.86 & 2.40 & 52.00 & 100.00 \\
pct\_occupation\_social\_service & 9056 & 1.73 & 2.14 & 0.00 & 100.00 & pct\_heating\_fuel\_wood & 9042 & 1.56 & 5.74 & 0.00 & 69.70 \\
pct\_occupation\_legal & 9056 & 1.29 & 2.01 & 0.00 & 55.56 & pct\_heating\_fuel\_none & 9042 & 3.43 & 5.36 & 0.00 & 49.40 \\
wfir\_risk\_score & 9106 & 1.08 & 1.54 & 0.00 & 5.00 &  &  &  &  &  &  \\
\bottomrule
\end{tabular}%
}
\end{table*}

\begin{table*}[t]
\centering
\caption{Summary statistics of census-tract-level demographic features in Florida.}
\label{tab:florida-tract-feature-summary}
\resizebox{\textwidth}{!}{%
\begin{tabular}{@{}lrrrr@{\hspace{1.2em}}lrrrr@{}}
\toprule
Feature & Mean & Std. & Min & Max & Feature & Mean & Std. & Min & Max \\
\midrule
urban\_rate & 90.29 & 26.02 & 0.0 & 100.0 & pct\_occupation\_education & 4.95 & 3.58 & 0.0 & 31.4 \\
median\_household\_income & 79311.33 & 32930.30 & 7969.0 & 250001.0 & pct\_occupation\_healthcare\_support & 2.95 & 3.10 & 0.0 & 28.3 \\
mean\_household\_income & 105396.77 & 51053.99 & 13193.4 & 714087.8 & pct\_occupation\_protective & 2.33 & 2.60 & 0.0 & 27.1 \\
pct\_black\_alone & 14.31 & 18.65 & 0.0 & 98.6 & pct\_occupation\_food\_service & 5.83 & 4.51 & 0.0 & 60.0 \\
pct\_hispanic\_latino & 26.66 & 24.50 & 0.0 & 100.0 & pct\_commute\_carpool & 8.97 & 6.42 & 0.0 & 56.0 \\
pct\_white\_not\_hispanic & 52.41 & 28.48 & 0.0 & 100.0 & pct\_commute\_public\_transit & 1.30 & 2.98 & 0.0 & 50.0 \\
pct\_asian\_alone & 2.64 & 3.43 & 0.0 & 35.0 & pct\_commute\_walk & 1.59 & 3.42 & 0.0 & 61.3 \\
fem\_pct & 50.80 & 5.41 & 0.0 & 73.4 & pct\_complete\_plumbing & 99.06 & 1.94 & 75.2 & 100.0 \\
male\_pct & 49.20 & 5.41 & 26.6 & 100.0 & occupants\_per\_room & 3.52 & 4.66 & 0.0 & 100.0 \\
pct\_hs\_grad\_higher & 89.50 & 8.61 & 0.0 & 100.0 & pct\_no\_telephone & 0.96 & 1.88 & 0.0 & 61.1 \\
less\_9th\_pct & 4.40 & 5.04 & 0.0 & 48.4 & pct\_units\_1\_attached & 6.62 & 8.55 & 0.0 & 80.7 \\
pct\_bachelor\_higher & 33.72 & 17.27 & 0.0 & 95.7 & pct\_units\_3\_4 & 3.83 & 5.50 & 0.0 & 66.7 \\
unemployment\_rate & 4.92 & 3.99 & 0.0 & 40.0 & median\_rooms & 5.22 & 0.96 & 2.1 & 10.0 \\
poverty\_rate & 13.31 & 9.87 & 0.0 & 100.0 & pct\_occupied\_housing & 86.23 & 12.40 & 2.4 & 100.0 \\
eld\_65p\_pct & 23.43 & 14.98 & 0.0 & 91.9 & pct\_industry\_construction & 8.36 & 5.89 & 0.0 & 81.5 \\
youth\_18l\_pct & 18.21 & 7.82 & 0.0 & 48.1 & pct\_rent\_under\_500 & 3.09 & 7.83 & 0.0 & 100.0 \\
avg\_household\_size & 2.51 & 0.52 & 1.1 & 4.8 & pct\_rent\_500\_999 & 11.25 & 15.44 & 0.0 & 100.0 \\
pct\_owner\_occupied & 67.48 & 22.69 & 0.0 & 100.0 & pct\_rent\_1000\_1499 & 22.90 & 19.03 & 0.0 & 100.0 \\
pct\_foreign\_born & 21.62 & 18.54 & 0.0 & 89.8 & pct\_rent\_1500\_1999 & 24.07 & 18.39 & 0.0 & 100.0 \\
per\_capita\_income & 44227.02 & 24340.33 & 1606.0 & 273846.0 & pct\_rent\_2000\_more & 31.38 & 27.13 & 0.0 & 100.0 \\
median\_age & 45.03 & 11.02 & 19.2 & 81.5 & pct\_heating\_fuel\_oil & 0.14 & 0.59 & 0.0 & 15.4 \\
total\_population & 4411.75 & 2244.20 & 3.0 & 25856.0 & households\_cohabiting & 7.53 & 4.96 & 0.0 & 100.0 \\
built\_2014\_pct & 2.68 & 5.30 & 0.0 & 100.0 & pct\_households\_living\_alone & 28.55 & 12.88 & 0.0 & 100.0 \\
pct\_native\_alone & 0.37 & 1.34 & 0.0 & 33.3 & pct\_married\_now & 48.74 & 12.94 & 0.0 & 88.1 \\
pct\_occupation\_management & 11.89 & 6.96 & 0.0 & 76.1 & pct\_graduate\_degree & 12.85 & 9.22 & 0.0 & 75.2 \\
pct\_occupation\_business & 5.90 & 4.45 & 0.0 & 100.0 & pct\_units\_1\_detached & 55.75 & 28.59 & 0.0 & 100.0 \\
pct\_occupation\_computer & 2.92 & 2.82 & 0.0 & 26.9 & pct\_units\_2 & 2.17 & 3.74 & 0.0 & 63.6 \\
pct\_occupation\_architecture & 1.67 & 2.15 & 0.0 & 36.2 & pct\_complete\_kitchen & 98.59 & 2.32 & 72.0 & 100.0 \\
pct\_occupation\_social\_service & 1.46 & 1.91 & 0.0 & 20.7 & pct\_heating\_fuel\_wood & 0.12 & 0.60 & 0.0 & 20.5 \\
pct\_occupation\_legal & 1.41 & 2.14 & 0.0 & 27.9 & pct\_heating\_fuel\_none & 2.58 & 2.90 & 0.0 & 28.9 \\
hurricane\_risk\_score & 4.56 & 0.28 & 0.0 & 5.0 &  &  &  &  &  \\
\bottomrule
\end{tabular}%
}
\end{table*}

\subsection*{Annotation and model selection}

Tweets for the CA and FL cases were collected from Brandwatch using the following case-specific criteria:

\textbf{CA case.}
\begin{itemize}
    \item Search terms: \textit{wildfire}, \textit{bushfire}, \textit{California fire}, \textit{CAfire}, \textit{CALfire}
    \item Search location: California
    \item Search period: January 1, 2025 -- January 31, 2025
\end{itemize}

\textbf{FL case.}
\begin{itemize}
\item Search terms: \textit{hurricane helene}, \textit{hurricane helen}, \textit{tropical storm}, \textit{tropical cyclone}, \textit{hurricane}
\item Search location: Florida
\item Search period: September 20, 2024 -- October 5, 2024
\end{itemize}


To establish a ground-truth dataset for evaluating sentiment prediction on wildfire-related tweets, we applied a consistent manual annotation protocol to the CA and FL cases. This shared framework provided a common basis for comparing the performance of large language models (LLMs) across the two cases.

\textbf{Annotation framework.} For each case, two trained annotators labeled 2,000 tweets to classify sentiment as \textit{positive} (1), \textit{neutral} (0), or \textit{negative} (-1). Negative sentiment primarily reflected emotional distress such as fear, grief, or anger, which was often linked to property loss, evacuation, or emergency delays. Positive sentiment captured support, gratitude, or community solidarity. Neutral sentiment was assigned to tweets lacking affective tone or conveying factual information. To evaluate annotation consistency, we measured inter-annotator reliability using \textbf{Krippendorff’s Alpha} ($\alpha$), a robust coefficient designed for nominal data and multiple coders:

\begin{equation} 
\alpha = 1 - \frac{D_o}{D_e} \label{eq:kripp_alpha} \end{equation}
where $D_o$ represents the \textit{observed disagreement} among annotators and 
$D_e$ represents the \textit{expected disagreement} under conditions of random annotation. The observed disagreement is computed as 
$D_o = \frac{1}{N} \sum_{i=1}^{N} \delta(a_{i1}, a_{i2})$, 
where $N$ is the total number of annotated items and $a_{i1}$ and $a_{i2}$ denote the sentiment labels assigned by the two annotators for item $i$. The function $\delta(a,b)=1$ when the annotations differ and $0$ when they agree. The expected disagreement is defined as 
$D_e = \sum_{c_1 \ne c_2} p(c_1)\, p(c_2)$,
where $p(c)$ is the proportion of annotations assigned to category $c$.

A higher value of $\alpha$ indicates stronger agreement after adjusting for chance, while values near zero suggest agreement consistent with random labeling. The final reliability score reflects substantial consistency and interpretive alignment among annotators, ensuring that the labeled dataset provides a reliable foundation for model evaluation.

\textbf{LLM evaluation.} We evaluated a suite of eight LLMs, including open-source models deployable on local GPUs (e.g., \texttt{Mistral-8B}, \texttt{DeepSeek-8B}, \texttt{Llama-8B}) as well as proprietary API-based systems (e.g., \texttt{GPT-4o},\texttt{GPT-5}, \texttt{Gemini~2.5}). Each model was assessed using a standardized three-step prompting workflow to ensure consistency across evaluations:

\begin{itemize}
    \item Location detection -- further identifying tweets originating from or referencing the study region.
    \item Event relevance identification --further filtering messages to retain only wildfire-related content.
    \item Sentiment classification -- assigning each tweet to one of three sentiment categories.
\end{itemize}

All models were evaluated under both zero-shot and few-shot configurations. To preserve the authenticity and inherent noise of the social media data, only minimal preprocessing was applied to the collected tweets.

\begin{table}[h!]
\centering
\caption{Model performance summary}
\label{tab:model_summary}
\begin{tabular}{lcc}
\toprule
\textbf{Model} & \textbf{Accuracy} & \textbf{F1} \\
\midrule
Mistral-8B-Instruct-2410        & 75.4\% &0.76 \\ 
DeepSeek-R1-Distill-Llama-8B    & 74.7\% & 0.75 \\ 
Gemini 2.5                      & 65.0\% & 0.469 \\
GPT-5 Instant                   & 60.0\% & 0.600 \\ 
GPT-4o-mini                     & 60.0\% & 0.413 \\
GPT-4o                          & 55.0\% & 0.393 \\
GPT-5 Thinking                  & 55.0\% & 0.413 \\
LLaVA-v1.6-Vicuna-7B            & 55.5\% & 0.53 \\ 
\bottomrule
\end{tabular}
\end{table}

Based on Table \ref{tab:model_summary}, although multimodal models such as \texttt{LLaVA} demonstrated versatility, they performed less effectively on this text-only dataset. Among all evaluated systems, \texttt{Mistral-8B} achieved the highest and most stable accuracy, closely followed by \texttt{DeepSeek}. We therefore selected \texttt{Mistral-8B} as the baseline sentiment classifier for downstream analysis due to its strong performance, efficient inference on A100 GPUs, and architecture that is well suited for fine-tuning under constrained computational settings. 

Importantly, our evaluation also reveals a broader insight: while tested LLMs achieve competitive accuracy, they nonetheless exhibit systematic biases in sentiment interpretation—particularly when processing emotionally nuanced, community-specific, or geographically contextualized language. These biases underscore the need for fairness-aware modeling approaches such as ours, which explicitly incorporate demographic and spatial diversity into disaster communication analysis.

\end{document}